# Formulation of an equivalent GNF model as an efficient approximation for flow of polymer solutions described by FENE-P

Anirban Ghosh, Raghav Kumar, Indranil Saha Dalal[a]

**AFFLIATIONS**

Department of Chemical Engineering, Indian Institute of Technology, Kanpur 208016, India

[a]Author to whom correspondence should be addressed: indrasd@iitk.ac.in

**ABSTRACT**

The molecular constitutive models, like FENE-P for polymer solutions, are known to have convergence issues at relatively larger flow rates. In this study, we investigate the possibility of a numerically efficient GNF-based approximation of FENE-P, which would closely approximate the flow field. Firstly, we compare the flow fields predicted by FENE-P and an equivalent GNF model. For these studies, we considered the flow around a sphere and selected the Carreau-Yasuda model as the representative GNF. This is made equivalent to the FENE-P by selecting parameters to equalize the viscosity-shear rate dependence. Our results show severe deficiencies of the GNF model, owing to its inability to account for chain stretching, particularly near the stagnation points. Next, the effect of extensional components on the local viscosity was added by formulating an equivalent GNF-X [*Journal of Rheology 64, 493 (2020)*] model. Even this failed to capture the asymmetry in the stress and flow profiles and predicted very large stresses at both stagnation points, relative to FENE-P. Hence, we proposed a novel modified formalism (denoted as GNF-XM) that was able to capture all trends successfully. The drag coefficients from GNF-XM agreed well with FENE-P predictions for all flow rates considered. Significantly, the computational times required to solve the flow field with GNF-XM is about an order of magnitude lower than that of FENE-P, especially at higher flow rates. Thus, we have successfully formulated a highly efficient GNF-based approximation to the FENE-P, whose formalism can be extended to other similarly complicated constitutive models.

## 1. Introduction

In our everyday life we encounter and interact with fluids almost every second. Starting from the air we breathe, the water we drink, to the blood which flows in our body, all are examples of fluids. Analysis of the mechanics of any fluid involves a constitutive equation, which connects the stress to the strain rate imposed upon the fluid. The earliest constitutive model uses a linear relation between the two, with the viscosity being the proportionality constant. These are called Newtonian fluids and are satisfactory models for the typical fluids surrounding us, like water, air, honey and alcohol. Mathematically this is given by Eqn. (1).

$$\vec{\vec{\tau}} = \eta \vec{\vec{\dot{\gamma}}} \qquad (1)$$

Where $\vec{\vec{\tau}}$ is the applied stress on the fluid, $\vec{\vec{\dot{\gamma}}}$ is the strain rate and $\eta$ is the viscosity of the fluid. Although this law explained the flow behaviour for many of the fluids, it was not universal. Later on, scientists discovered many fluids that did not follow Eq. (1). This entire class of fluids were termed as non-Newtonian, since they did not follow the Newtonian constitutive model (Eq. (1)). The deviation from the Newtonian behaviour can be in different ways, like decrease or increase in viscosity with increase in shear rate, change in viscosity with time, or even deforming slightly over some period of time on removal of stress, each of which represents a different class of constitutive equations. Knowledge and investigation of such fluids is important, since most fluids that we encounter in our daily lives are non-Newtonian. This list is enormous, and we attempt to provide a few notable examples. The most important non-Newtonian biological fluid is blood, which enables the transport of materials in our body. Apart from this, the fluids in the cell cytoplasm, the environment that a cell typically encounters and even a suspension of cells or colloidal particles is non-Newtonian. As we move away from biology, the modern-day industry is primarily concerned with the processing of non-Newtonian fluids – polymer solutions, polymer melts, paints, toothpaste, gels etc. Experiments on non-Newtonian fluids have been performed ever since the era of Poiseuille, who formed his famous equation while studying blood flow.

In recent times, owing to the huge progress in CPU processing power, theoretical models and computer simulations have become equally important. Researchers have developed and used constitutive models and simulation algorithms for a variety of complex (or non-Newtonian) fluids. Such developments for a wide variety of complex fluids are summarized elsewhere [1, 2]. In recent times, computer simulations have contributed to a better understanding for different flow situations. These include internal flows i.e., flows inside boundaries, like flow through pipes and channels, as well as external flows around an object. The latter includes flow around particles of various shape, size and orientation – like flow around cylinders, spheres and ellipse. However, all these complex fluids have their own constitutive models, which should ideally be developed individually for each case, starting right from the changes induced in the microstructure due to imposed flow fields.

In this study, we have focused primarily on dilute polymer solutions. For these, the Generalized Newtonian Fluid (GNF) class of models were the earliest. These involved simple corrections to the viscosity (as a function of the local shear rate), leading to the power-law and Carreau-Yasuda models [2]. Such viscosity corrections are simple to implement within existing CFD techniques. Thus, such GNF models are incorporated within a variety of CFD packages and have been widely used by researchers over decades for various types of problems, even recently [3-6]. From various experiments performed by Weissenberg [7], Leslie and Tanner [8] and other rheologists over last few decades, it is known that the behavior of dilute polymer solutions can't be completely explained by simple shear-thinning fluid models. Polymer solutions are viscoelastic in nature and show shear-thinning, memory and elastic effects. Some famous phenomena that exhibit these are the Weissenberg rod-climbing effect, die-swell effect and flow of silly putty over a long time [2]. However, it is known that the GNF models, which are still highly popular, capture only the shear thinning effects. Thus, there were further developments, leading to the development of differential equations as constitutive models, like the FENE-P, formulated by Bird and coworkers [9]. The FENE-P exhibits shear thinning, memory and elastic effects and is incorporated in some CFD packages. It attempts to model a polymer solution by representing the polymer chain as a dumbbell connected by a FENE spring. The solvent around the chain is modeled by a continuum (a Newtonian fluid). A related, but earlier model, the Oldroyd-B [10], uses a Hookean spring. Thus, from the fundamental viewpoint, the FENE-P is much closer to reality than the GNF models. The physics is better than Oldroyd-B as well, since a polymer chain is finitely extensible, an aspect not captured by the Hookean spring used in Oldroyd-B.

Although the FENE-P is much more accurate and has been around for four decades, it has not been used by researchers as extensively as the other simpler and relatively inaccurate models, like the power-law and Carreau-Yasuda. In fact, this model has been rarely used for even reasonably complicated flow situations. This is partly due to the additional complexities encountered while using the FENE-P in CFD simulations. The FENE-P model has differential equations for each stress component, thereby increasing the set of equations that need to be solved. Solving such a set of differential equations for even visibly simpler flow problems is not trivial, and often require highly specialized algorithms. Such techniques to solve for viscoelastic constitutive models like the FENE-P, for various flow problems, are summarized in a recent review article [11]. However, we note that, even with such improvements of algorithms, it is extremely difficult to obtain converged results with the FENE-P for a wide variety of flow situations. Thus, even though the FENE-P has been used recently for some benchmark problems like the flow around a sphere [12] and cylinder [13], the computations are limited to low Weissenberg numbers (which we define later), mostly well below 10. Hence, for the constitutive models for polymer solutions, the FENE-P lies at one end, which incorporates most of the physics but is nearly impossible to compute and converge even for moderately large flow rates. At the other end, we have the GNF models, which are much easier to implement and converge, for a wide variety of flows. The latter ones, or models like these, are highly preferable from a practical standpoint. This is particularly true if one is interested to obtain an approximate solution to a flow situation which is reasonably close to the actual, i.e. which would have been obtained if the FENE-P could have been successfully solved for the same problem. Thus, a natural question that arises - how different will be the two solutions – obtained from the FENE-P and the GNF models? If the predictions are significantly different, the next immediate task can be to bring the solutions closer by some further adjustments to the GNF model.

In this study, we explore both these aspects for a classic flow situation – the flow of a polymer solution around a sphere. First, we ascertain the difference in the solutions obtained from the FENE-P and the GNF model for same flow rates. For this comparison, the GNF model must be made "equivalent" to the FENE-P model in some sense. In this study, for a known set of parameters of the FENE-P model, an equivalent GNF model is constructed with the constraint that the viscosity-shear rate dependence is the same. This leads to an equivalent Carreau-Yasuda model, as we show later in this manuscript. Our study shows that the predicted stress profiles from the equivalent GNF model are significantly different, even qualitatively, for the flow rates considered here. However, the predictions agree very well for an internal flow scenario (flow in a tube for this study).

These observations may be expected, since the flow in a tube is purely shearing in nature, and the GNF model captures the shear thinning effect. Since the shear thinning behavior is matched for the construction of the equivalent GNF model, the predictions can be expected to be similar for a purely shearing flow. However, the flow around a sphere would have local extensional components, in addition to shear, at various points in the domain. Since the GNF model used no extensional flow information for its parameterization, its failure can be expected. However, since we need an approximate and reasonable solution, can the GNF be "improved" if it was constructed such that the local viscosity adjusts depending on the proportion of local extensional and shearing components in the flow field? In the second part of this manuscript, we primarily focus on this aspect. The equivalent GNF model is constructed in such a manner that its viscosity depends on the magnitudes of local extensional and shearing components. To achieve this, we have used the recently proposed GNF-X formalism [14]. The viscosity for various extensional and shear rates are given by the respective steady state values for the FENE-P model. To the best of our knowledge, the GNF-X formalism has not been used earlier for any external flow simulation involving polymer solutions. Our results indicate that, with some suitable modifications, such a formulation does manage to provide a GNF model that shows similar predictions as FENE-P.

Thus, in this manuscript, we present a way to obtain an equivalent GNF model that can provide an approximate solution which is reasonably close to that obtained from the FENE-P. If such a model can exist, one can obtain solutions to complicated flow scenarios without facing issues related to convergence, like the FENE-P. The article is organized in the following manner. In the methodology section, we discuss in details about the different constitutive equations used in this study as well as the simulation setup. Next, we show our comparison of the equivalent GNF and FENE-P, followed by comparisons with the GNF-X model. Then we propose our modifications to the GNF-X model and show the improvements over the original GNF-X. Throughout in this study, velocity and stress contours and drag coefficients are used for comparisons. Finally, we also show that the resulting modified GNF-X model is computationally cheaper by an order of magnitude than the FENE-P, especially as flow rates are increased.

## 2. Methodology

2.1 Governing Equations

For flow simulations of an incompressible fluid, we solve the continuity and momentum balance equation, along with the appropriate constitutive model. The continuity and momentum balance equations for any incompressible fluid are given as:

$$\nabla . \vec{V} = 0 \tag{2}$$

$$\rho(\vec{V} . \nabla \vec{V}) = \nabla . \vec{\sigma} \tag{3}$$

$$\vec{\sigma} = -p\vec{I} + \vec{\tau} \tag{4}$$

Here, $\vec{V}$ is the velocity field, $\rho$ denotes the fluid density, and $\vec{\sigma}$ is the total stress tensor. The stress tensor is a sum of an isotropic part, given by the pressure ($p\vec{I}$) and a deviatoric part ($\vec{\tau}$). The deviatoric stress tensor is a function of the deformation of the material and is estimated with the help of a constitutive relation. As discussed before, such constitutive equations have been developed over several decades, ranging widely in terms of included physics and complexity, even for the same material (polymer solution in this study). Here, we list the models that have been used in this study. Note that, for solving any flow problem, Eqns. (2)-(4) need to be used in addition to these constitutive models.

*2.1.1 Carreau-Yasuda (CY) model*

This is one of the most widely used GNF (Generalized Newtonian Fluid) model for various applications, especially for polymer solutions. Being a GNF model, the viscosity is specified as a function of the shear rate. The constitutive equation is given as follows [15,16]:

$$\vec{\tau} = \eta(\dot{\gamma})\vec{D} \tag{5}$$

$$\eta(\dot{\gamma}) = \eta_\infty + (\eta_0 - \eta_\infty)(1 + (\lambda\dot{\gamma})^a)^{\frac{(n-1)}{a}} \tag{6}$$

Here $\eta_0$ is the zero-shear rate viscosity, $\vec{D}$ is the rate of strain tensor, $\eta_\infty$ is the limiting viscosity at very high shear rates, $\lambda$ denotes the relaxation time, and $n$ and $a$ are the characteristic values for a particular fluid. These values ($n$, $a$ and $\lambda$) are normally obtained by fitting this function to the experimental data of viscosity with the shear rate for given $\eta_\infty$ and $\eta_0$, which may be known beforehand for a polymer solution. Note, Eqn. (5) is true for any GNF fluids, while Eqn. (6) is specific to the CY model. Also, note that, this can be considered as an improvement to the power-law model, which is able to show a plateau in the viscosity for weak and extremely high shear rates, and a power law in the intermediate regime. In this study, the CY model is used as the GNF representation, whenever needed.

### 2.1.2 Extended Generalized Newtonian Fluid Model (GNF-X)

This model represents a recent correction to the existing GNF framework [14]. GNF models have been particularly successful in capturing the effects due to shear thinning of polymeric fluids. The shear-thinning effect is incorporated in the viscosity-shear rate relationship for the GNF models. However, for various applications, the local flow field is almost never a simple shear and generally consists of both extensional (stretching) and shearing components. The contribution due to the extensional part is not captured by the typical GNF model, thus requiring an extension. The GNF-X model has contributions to the local viscosity from both extensional and shear parts of the local flow field with appropriate weights. This is given by Eqns. (7) - (10):

$$\overleftrightarrow{\tau} = \eta(\dot{\gamma})\vec{D} \tag{7}$$

$$\eta(\dot{\gamma}) = (1-W)\eta_s(\dot{\gamma}'_S) + W\eta_E(\dot{\gamma}'_E) \tag{8}$$

$$\text{where } W = \frac{\dot{\gamma}'^2_E}{\dot{\gamma}^2} \tag{9}$$

$$\text{and } \dot{\gamma}^2 = \dot{\gamma}'^2_S + \dot{\gamma}'^2_E \tag{10}$$

Here, $\eta_s$ and $\eta_E$ are the contributions to the local viscosity from the shear and extensional parts, respectively. $\dot{\gamma}'_S$ and $\dot{\gamma}'_E$ are the shear and extensional components, respectively, of the total strain rate $\dot{\gamma}$. Any appropriate GNF model can be used to estimate the shear contribution to the viscosity ($\eta_s$) as a function of the shear strain ($\dot{\gamma}'_S$), which we refer to as simply the shear viscosity. In our study, we have used the Carreau-Yasuda model to estimate the shear viscosity within GNF-X. As discussed later, this separation of the total strain rate into its shear and extensional parts is performed by transforming the coordinates to the streamline coordinate system.

### 2.1.3 Oldroyd-B Model

Oldroyd-B model belongs to the class of molecular models, which are derived by using some aspects of the structure and dynamics of the material (i.e. polymer solution for this study) at the microscale. This model represents polymer chains by Hookean dumbbells (two beads connected by a spring) that are immersed in a Newtonian solvent. Phenomenologically, this can also be derived by taking a Hookean spring in parallel with a Newtonian dashpot, in series with another Newtonian dashpot [2, 10]. Thus, the deviatoric stress can be written as a sum of the solvent ($\overleftrightarrow{\tau}_s$) and polymeric ($\overleftrightarrow{\tau}_p$) contributions, as follows:

$$\overleftrightarrow{\tau} = \overleftrightarrow{\tau}_S + \overleftrightarrow{\tau}_P \tag{11}$$

Here, the solvent is assumed to be a Newtonian fluid. Thus, the solvent stress is given by Eqn. (1). The polymeric stress is related to the strain rate by the following relation:

$$\overleftrightarrow{\tau}_P + \lambda \stackrel{\nabla}{\overleftrightarrow{\tau}_P} = 2\eta_P \vec{D} \tag{12}$$

Where $\lambda$ is the relaxation time and $\eta_P$ is the polymeric contribution to the zero-shear rate viscosity. The $\nabla$ symbol on $\vec{\tau}_P$ denotes the upper convected derivative. The upper convected derivative of a tensor $\vec{A}$ is given by Eqn. (13):

$$\overset{\nabla}{\vec{A}} = \frac{D\vec{A}}{Dt} - (\nabla \vec{V}^T . \vec{A} + \vec{A} . \nabla \vec{V}) = \frac{\partial \vec{A}}{\partial t} + \vec{V}.\nabla \vec{A} - (\nabla \vec{V}^T . \vec{A} + \vec{A} . \nabla \vec{V}) \qquad (13)$$

### 2.1.4 FENE-P Model

Physically, this is quite similar to the Oldroyd-B model and belongs to the class of molecular models. Here, the polymer chain is mimicked by a single finitely extensible spring (instead of Hookean as in Oldroyd-B) spring (with two beads), immersed in a Newtonian solvent. The microscopic configurational state of the polymer chain is given by the conformation tensor ($\vec{A}$). This is the dyadic product of the polymer connector vector ($\vec{R}$) with itself. The connection of the polymeric stress and the conformation tensor is given by [17]:

$$\vec{\tau}_P = -\eta_P \overset{\nabla}{\vec{A}} = \frac{\eta_P(f(\vec{A})\vec{A} - a\vec{I})}{\lambda} \qquad (14)$$

where 
$$a = \frac{1}{1 - \frac{3}{L^2}} \qquad (15)$$

The evolution equation for the conformation tensor is given by:

$$\overset{\nabla}{\vec{A}} = -\left( \frac{f(\vec{A})\vec{A} - a\vec{I}}{\lambda} \right) \qquad (16)$$

Here $\eta_P$ is the polymeric contribution to the zero-shear rate viscosity, $\lambda$ is the relaxation time and $L^2$ is the extensibility parameter of the polymer chain.

### 2.2 Simulation Schematic

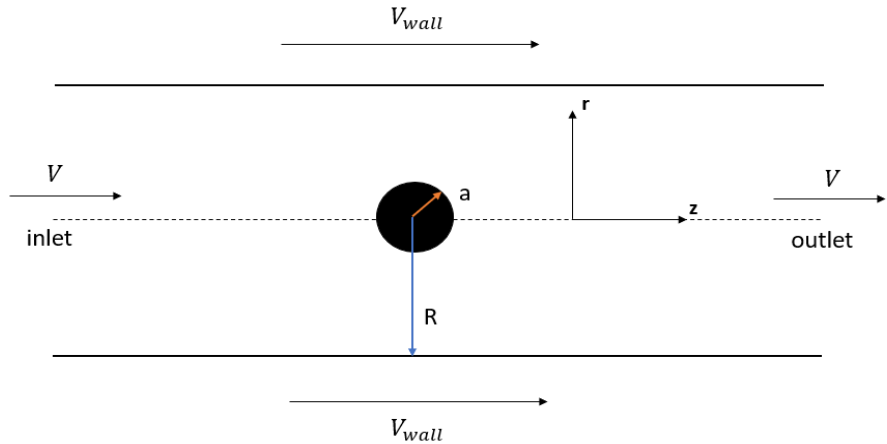

**FIGURE 1**: A schematic of the simulation setup of the problem. Since the flow is axisymmetric in nature, the r-z coordinate is used for the simulations. The coordinate system along with the boundary conditions are shown in the schematic.

In this manuscript, we study the settling of a small sphere in a polymer solution. Thus, for our simulations, we have taken an outer cylinder (as the bounding domain) with a sphere at the center of the cylinder's body. Since the system is symmetric about the axis of the cylinder, we can solve it as a 2D axisymmetric problem. For studies using the GNF models, the radii of the sphere and the bounding cylinder are taken as 0.121 m and 1 m, respectively, to create an unconfined flow around the sphere. The cylinder has been kept sufficiently long in the flow direction to prevent any boundary effects of inlet and outlet on the flow around the sphere. In later sections while using the GNF-X model, the diameter of the sphere has been changed to obtain different values of the Weissenberg number (defined later in this manuscript) for a constant Reynolds number (Re). In all such cases, to create an unconfined flow, we have taken the radius of the cylinder to be 14 m and the length to be 42 m.

For our simulations, the following boundary conditions are imposed:

- Constant inlet velocity V
- Cylindrical wall velocity $V_{wall} = V$
- No-slip boundary condition at the sphere surface
- Ambient pressure at the outlet

Here, V is the uniform fluid velocity past the sphere along the z direction. ANSYS Polyflow was used for the comparative study of FENE-P and the equivalent GNF model, while COMSOL Multiphysics V5.6 was used for the comparisons of FENE-P and the equivalent GNF-X models. In all cases, we used the in-built meshing program in both ANSYS and COMSOL.

In our study, we used the velocity field and the stress contours to compare the predictions of various models. Additionally, drag force is used as a measure for the stress acting around the surface. The drag force is calculated by integrating all contributions in the flow direction from the stresses acting on the sphere. In our simulations, the flow is oriented along the z-direction. The magnitude of the drag force is calculated as follows:

$$F_d = \int (-p\vec{\vec{I}} + \overrightarrow{\tau_S} + \overrightarrow{\tau_P})\vec{n}.\vec{k}.dS \tag{17}$$

Here, $\overrightarrow{\tau_S}$ and $\overrightarrow{\tau_P}$ represent the solvent and polymeric contributions to the extra stress tensor, respectively, and $p$ represents the pressure. $\mathbf{k}$ denotes the unit vector in the flow direction and $\vec{n}$ is the outward normal for the area element $dS$.

For the flow of non-Newtonian fluids, the drag coefficient deviates from the value for Newtonian fluids. This deviation is captured by the drag correction coefficient, which is defined as:

$$\chi = \frac{K}{K_N} = \frac{F_d}{F_{newtonian}} = \frac{C_d(non-Newtonian)}{C_d(Newtonian)} \tag{18}$$

From our simulations, we calculate the variation of this ratio with the dimensionless quantity Wi (Weissenberg Number). This is defined as:

$$Wi = \frac{\lambda V}{L} \tag{19}$$

Here, $\lambda$ is the relaxation time of the polymer chain, $V$ is the characteristic velocity and $L$ is the characteristic length scale of the flow problem. In this study, $L$ is the diameter of the sphere ($D_s$). Also, all simulations performed in this study lie in the creeping flow regime. Thus, the Reynolds number considered is very small ($Re \ll 1$) for all simulations. The Reynolds number is defined as:

$$\text{Re} = \frac{\rho V D_s}{\eta} \tag{20}$$

Here, $\rho$ is the density of the fluid, $V$ is the velocity of fluid, $D_s$ is the characteristic length scale of the problem (which is the diameter of the sphere for our problem) and $\eta$ is the viscosity of the fluid.

2.3 Grid Independence Test

Table 1: Details of the mesh created for the grid-independence studies

| Meshes | Laminar flow (elements) | Viscoelastic flow (elements) |
|---|---|---|
| Mesh 1 | 3304 domain, 210 boundary | 3108 domain, 202 boundary |
| Mesh 2 | 6033 domain, 273 boundary | 5791 domain, 265 boundary |
| Mesh 3 | 11463 domain, 435 boundary | 10973 domain, 427 boundary |

Firstly, we perform the grid independence tests on the 2D axisymmetric geometry. Note that COMSOL Multiphysics produces meshes based on the physics of the problem. Thus, for COMSOL simulations, grid independence tests are performed for two different physics - laminar flow (for GNF-X) and viscoelastic flow (for FENE-P). Throughout, the COMSOL meshing module is used to create the mesh. We considered three different mesh sizes for the mesh independence tests. The number of elements for these different meshes are summarized in Table 1.

For simulations in ANSYS Polyflow, we have taken three different meshes with 4636, 10336, and 22422 elements. This software is used for simulations involving the FENE-P and GNF models, in the first half of the manuscript. The mesh is generated using the ANSYS meshing tool. A schematic of all meshes used for these simulations are shown in Fig. 2.

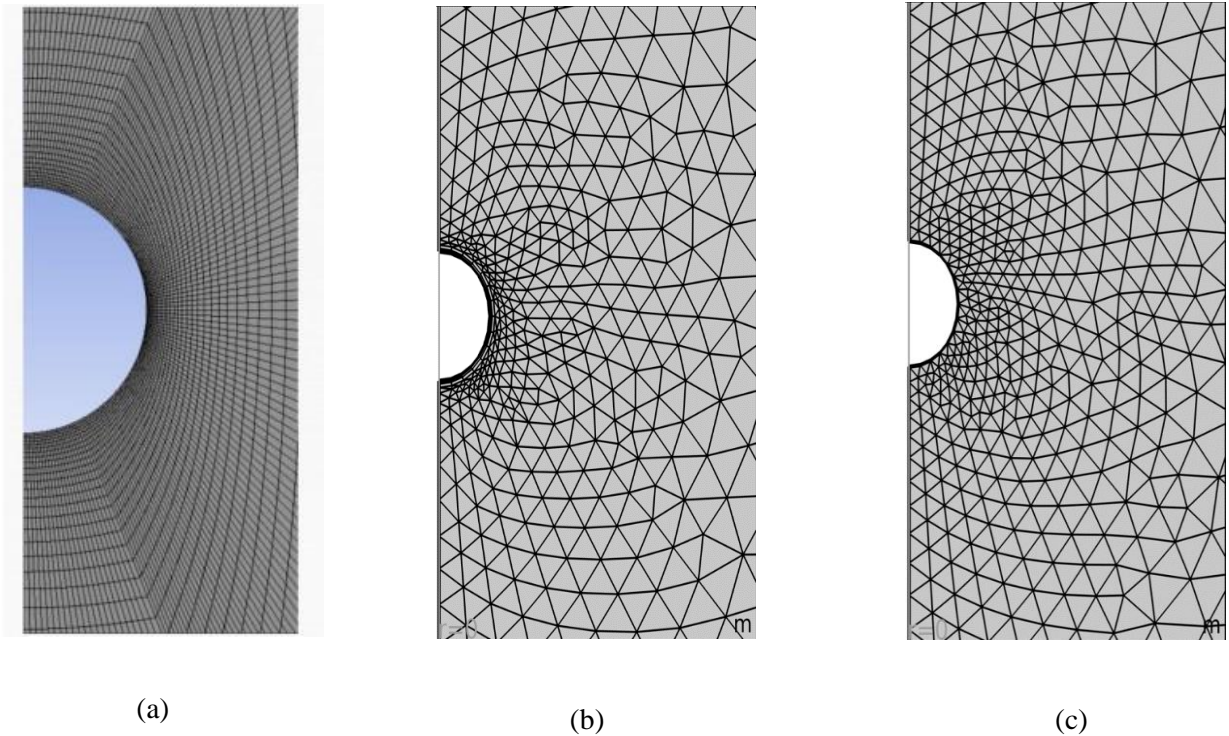

(a)  (b)  (c)

**FIGURE 2:** Schematic of the meshes used in simulations: (a) non-uniform quadrilateral meshes used for simulations in ANSYS Polyflow, (b) Mesh 2 (Table 1) used in grid independence tests for Newtonian fluid simulations in COMSOL and (c) Mesh 2 (Table 1) used in grid independence tests of the FENE-P model in COMSOL

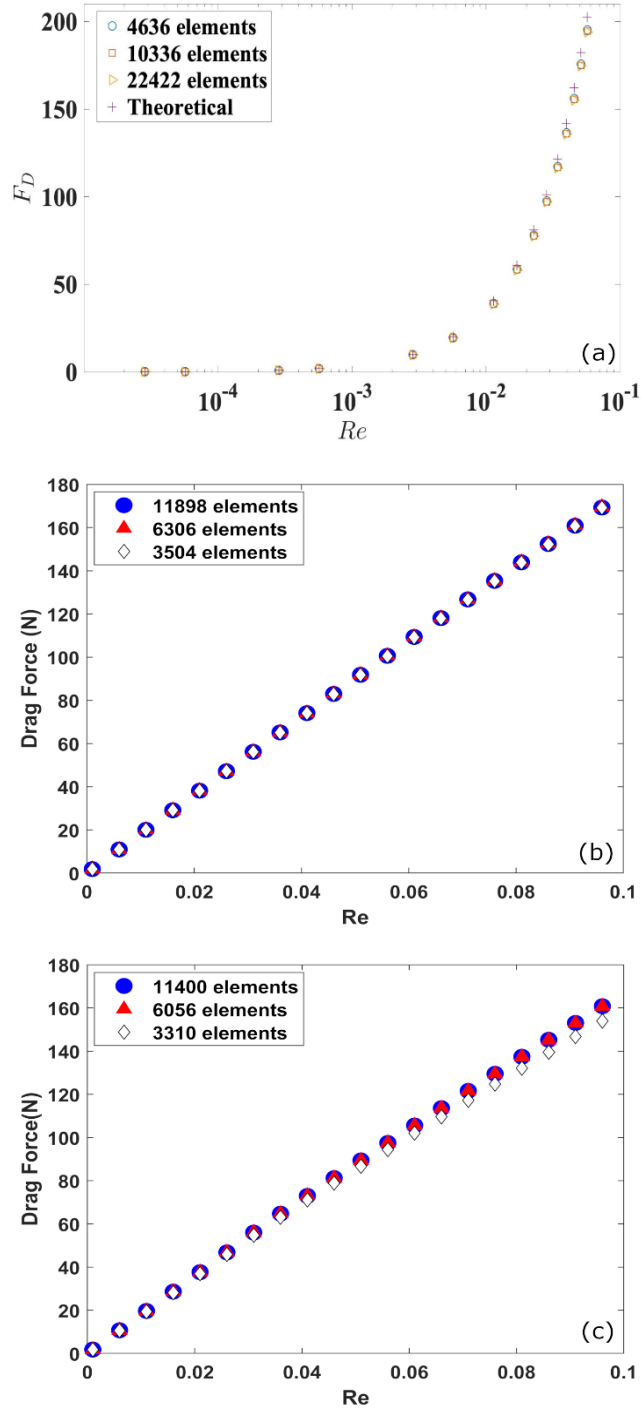

**FIGURE 3:** (a) Grid independence studies for (a) Newtonian fluid in ANSYS Polyflow, (b) Equivalent CY fluid for the FENE-P model used in mesh independence study in COMSOL Multiphysics and (c) FENE-P fluid in COMSOL Multiphysics with $\rho$ as 1 kg/m$^3$, $L^2$ as 500, relaxation time $\lambda$ as 0.749 s, zero shear viscosity $\eta_0$ as 13.76 Pa.s and solvent viscosity $\eta_s$ as 8.12 Pa.s

For the Newtonian fluid simulations, the density $\rho$ is 1 kg/m$^3$, viscosity $\eta$ is 13.76 Pa.s, and diameter of the sphere $D_s$ is 0.484m. For the CY GNF fluid, the zero shear viscosity $\eta_0$ is 13.76, $\eta_\infty$ is $10^{-4}$ Pa.s, and $a = 2$. Rest of the parameters were obtained by fitting the model to viscosity data obtained from the parameters used for the FENE-P as mentioned in the caption of Fig. 3.

The tests on the 2D geometry in ANSYS shows no improvement in the accuracy of results with an increase in the number of elements beyond 4636. So, these three grids of 4636, 10336, and 22422 elements predict almost identical values of drag. For our simulations, we select the grid with 10336 elements to maintain a balance between accuracy and computational cost.

Mesh independence test performed in COMSOL provide a similar trend. The three meshes predict nearly indistinguishable values of drag for the case of Newtonian fluid, as observed in Fig. 3(b). For the FENE-P fluid, a similarly good convergence of the meshes is noted. However, for FENE-P, the results for mesh 2 and 3 are closer relative to those for mesh 1, but the differences are small (less than 10%). This can be attributed to high elastic stresses, which are perhaps captured better by mesh 2 and 3, relative to mesh 1. However, owing to the computational complexity and the difficulty of convergence, we select mesh 2, which provides equivalent predictions as the finer mesh 3.

## 3. Results and Discussion

### 3.1 Comparison of FENE-P and CY-GNF

For our simulations, we used the following properties of the polymeric fluid. The rheological properties were taken from the study by Arigo et al. [18]

Table 2: Rheological Properties of the polymer solution used in our simulations

| | |
|---|---|
| $\rho$ (kg/m$^3$) | 1 |
| $\eta_0$ (Pa.s) | 13.76 |
| $\eta_s$ (Pa.s) | 8.12 |
| $\lambda$ (s) | 0.749 |

Firstly, we studied our problem using the Oldroyd-B constitutive equation, before FENE-P. Being a similar but slightly simpler model than the FENE-P, this would enable a better understanding of the trends. Our simulations with the Oldroyd-B model converged up to Wi = 3.2.

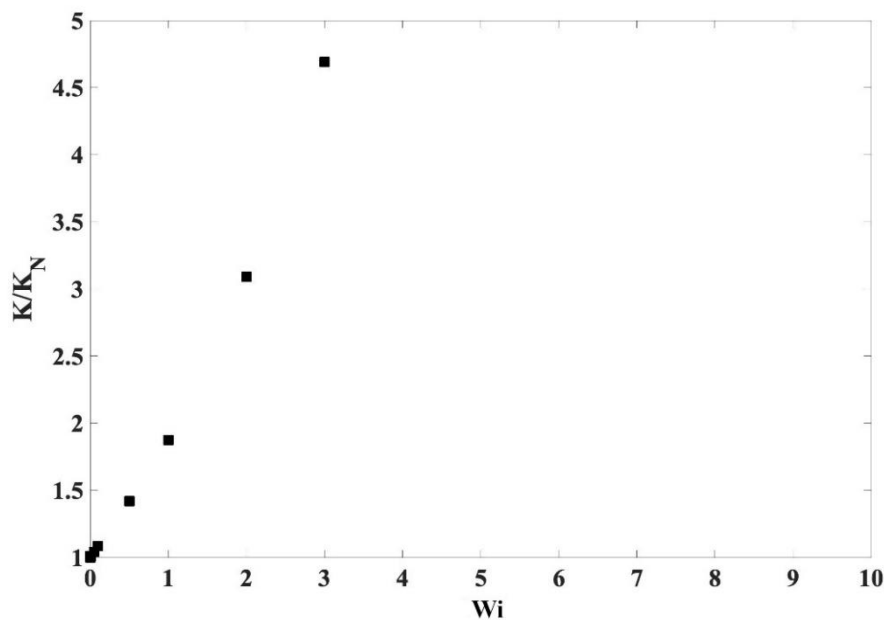

**FIGURE 4:** Variation of the drag correction coefficient with Wi for the Oldroyd-B model

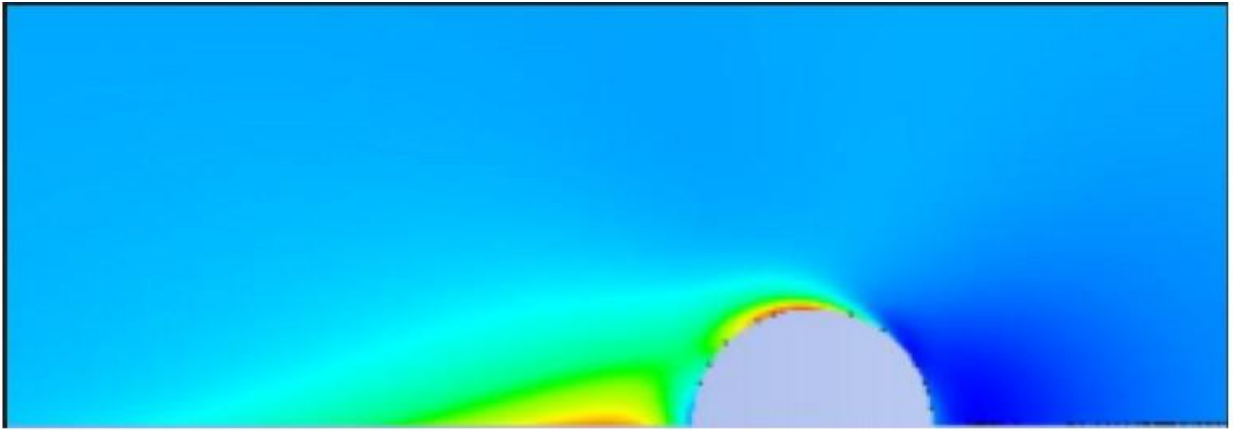

**FIGURE 5**: Normal stress contour for the Oldroyd-B model. The flow is from right to left. Notice the high stress region near the rear stagnation point.

The limitation on the maximum explorable Weissenberg number is the well-known high Weissenberg number problem. This is caused due to the infinite extensibility of the Hookean spring in the Oldroyd-B model. As the Wi is increased, a high normal stress zone is created in the wake of the sphere, which eventually diverges. This wake in stress can be observed in the contours presented in Fig. 5. This divergence eventually causes the simulations to fail. Fig. 4 shows the variation of the drag correction coefficient for the Oldroyd-b model. This increases with Wi in the range considered in our simulations. This is attributed to the increased normal stress as mentioned before, at the rear stagnation point, which causes the drag force to deviate significantly from that of the Newtonian fluid.

We expect similar trends with some differences for the FENE-P model, since it contains a finitely extensible spring. The parameters mentioned before are sufficient for the simulation of the Oldroyd-B model. However, due to the finite extensibility of the spring within the FENE-P model, we need an extensibility parameter ($L^2$) as well. Next, we compare the FENE-P model predictions with its equivalent GNF model. As discussed before, we will use the CY model as a representative of the GNF family. This comparison will show the inadequacies of the GNF class of models, leading to the need of the GNF-X. For the construction of the equivalent Carreau-Yasuda model (denoted as CY-GNF), we use the FENE-P model as the basis, with $L^2$ as 1.01, 1.1, and 2. For these FENE-P fluids, simulations of the simple shear flow are performed. These results provide the viscosity variation of the FENE-P fluid with shear rates (along with other material functions). These are shown in Fig. 6. Then, the corresponding parameters for the equivalent CY-GNF model are calculated by fitting this data to the viscosity-shear rate function used in the Carreau-Yasuda model (Eqn. (6)).

Table 3: Parameters of the equivalent CY-GNF models for the different FENE-P fluids (with different $L^2$ values) used in our study. The fits for the same are shown in Fig. 6.

| $L^2$ (dimensionless) | 1.01 | 1.1 | 2 |
|---|---|---|---|
| $\eta_\infty$ (Pa.s) | 0.0001 | 0.0001 | 0.0001 |
| $\eta_0$ (Pa.s) | 13.76 | 13.76 | 13.76 |
| $\lambda$ (s) | 2.7388 | -2.3752 | 0.5054 |
| n (dimensionless) | 0.3616 | 0.3621 | 0.3672 |

Figure 6 shows the material functions (as a function of shear rates) obtained with the FENE-P models (with different values of $L^2$). A correlation is observed between the material functions and $L^2$. With increasing $L^2$, both the shear thinning nature and the first normal stress difference (FNSD, shown in inset), at zero shear rate, increases. The equivalent CY-GNF model is obtained by fitting the viscosity-shear rate variation obtained from the various FENE-P models. The solid lines represent the CY-GNF fits to the respective FENE-P models (denoted by the $L^2$ value in the bracket in the legend). Since the CY model is a GNF, it can't predict FNSD for a simple shear flow. The parameter values obtained for the equivalent CY-GNF models are summarized in Table 3.

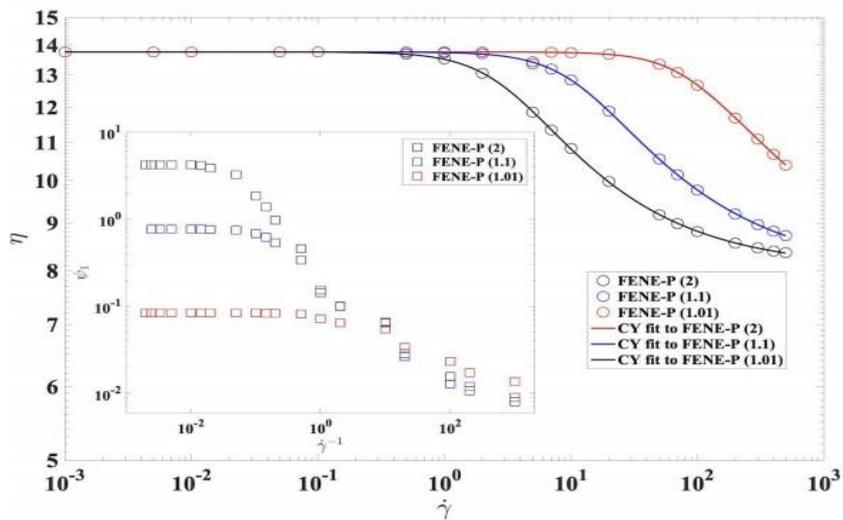

**FIGURE 6**: Variation of viscosity with shear rates for the FENE-P models (with varying $L^2$ values, as denoted in brackets). The solid lines show the corresponding fits for the equivalent CY-GNF models. The inset shows the variation of the first normal stress difference coefficient with shear rates, also obtained from simulations of the FENE-P fluids in simple shear flows.

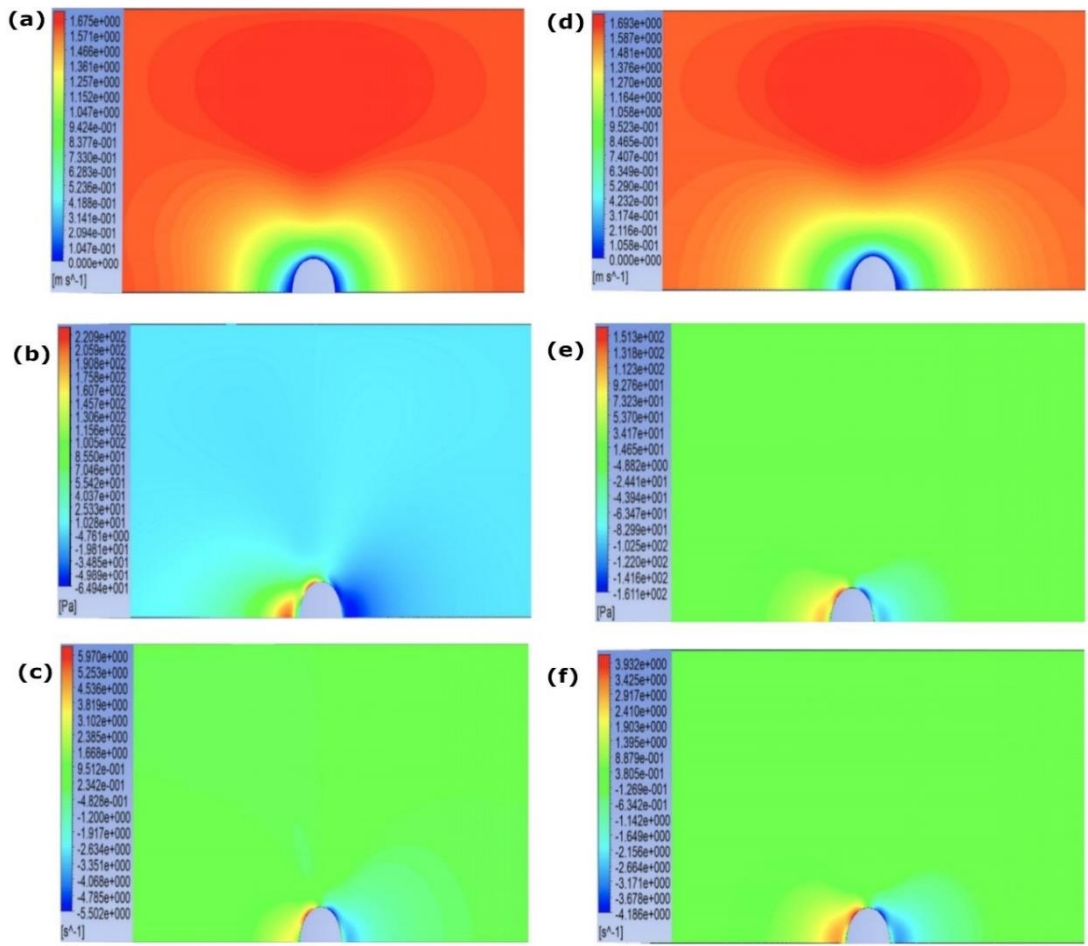

**FIGURE 7**. (a) Velocity contour (b) Stress contour in zz direction (c) degree of extension for a FENE-P fluid with $L^2 = 2$. (d), (e) and (f) are same as (a), (b) and (c) respectively, but for the equivalent CY-GNF model. Flow is from right to left, along the z-direction

Fig. 7 presents a comparison between the FENE-P model with $L^2 = 2$ and the equivalent CY-GNF, for flow around a sphere. From Figs. 7(a) and (d), the velocity contour is observed to be slightly more symmetric for the CY-GNF than the FENE-P fluid, where it is slightly compressed towards the sphere along the axis in the rear region. A much clearer asymmetry can be observed in the stress contours in Figs. 7 (b) and (e), which show the normal stresses in the flow direction. The FENE-P model generates a wake downstream of the sphere while the CY-GNF generates a near-symmetric normal stress contour. Furthermore, we calculated the individual components contributing to the drag force. It was observed that the pressure and shear stress contributions are of similar order of magnitude for either case. Thus, the entire increment in the drag force is due to the differences in the normal stress. As observed from Figs. 7(c) and (f), there is an increased degree of extension observed in FENE-P model, relative to the CY-GNF. Here, the degree of extension is calculated as $\frac{\partial w}{\partial z}$, which is essentially the z-component of the gradient of $w$ (velocity component in the z-direction). This higher extension causes an increase in the normal stress.

As mentioned before, the total stress $\vec{\vec{\tau}}$ is composed of a solvent stress $\vec{\vec{\tau}}_s$ and a polymeric stress $\vec{\vec{\tau}}_p$. As indicated in Fig. 8, the asymmetry of the stress in the FENE-P model (shown in Fig. 7(b)) likely originates from the polymeric contribution, while the solvent contribution, assumed to be Newtonian fluid, remains symmetric. Note, this is due to a largely similar asymmetry (as in Fig. 7(b)) observed in the polymeric stress contour shown in Fig. 8. To obtain further insights into this variation in the polymeric stress, we need to understand the effect of the flow field on the constituent particles (beads joined by FENE springs) at the microscopic length scales.

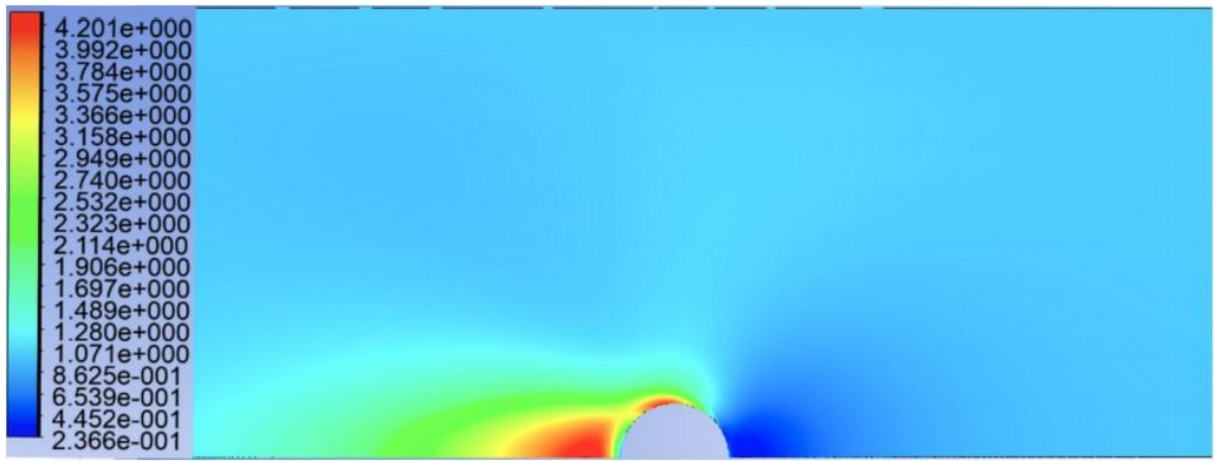

**FIGURE 8-** The variation of the zz-component of polymeric stress tensor for the FENE-P model with $L^2 = 2$. Flow is from right to left, along the z-direction

Most likely, the effect of the value of $\frac{dw}{dz}$ (Figs. 7 (c) and (f)) on the constituent polymer particles (represented by a dumbbell) is the origin of this asymmetry in polymeric stress. For this, let us imagine a pair of beads joined by a FENE spring approaching the sphere. Just upstream of the sphere, the velocity component $w$ of the bead closer to the sphere will be lower than the one further away, leading to a compression in the spring in the flow direction. Conversely, in the downstream region, the bead further away from the sphere moves away faster than the one closer to the sphere, leading to an extension in the spring. Under compression, the FENE spring remains in the Hookean regime (if initially it was near equilibrium i.e. in the Hookean regime) and doesn't diverge. On the other hand, under extension, the spring constant sharply diverges beyond a certain spring stretch (decided by the value of $L^2$), leading to high stress in the wake. It is highly likely that this asymmetry in the spring constant leads to the observed asymmetry in the polymeric stress. Thus, the results presented above suggest that the equivalent GNF model is most likely incapable of accurately predicting the flow behavior of polymeric fluids for the class of external flows (flow around a body). This is owing to the lack of elastic characteristics, which are inherently incorporated in the FENE-P model. Next, we probe if we can satisfactorily use an equivalent GNF model for any flow situation, instead of FENE-P.

Further, we compute the drag correction coefficients for the FENE-P and the equivalent GNF models. These are summarized in Fig. 9. Clearly, for the range of flow rates considered, the drag correction coefficient for the CY-GNF remains close to unity. However, those obtained for the FENE-P is non-monotonic and depends on the value of $L^2$. For some flow rates, the value obtained for the FENE-P model is about 40-50% higher than that of a Newtonian fluid. This discrepancy also depends strongly on the parameter $L^2$ and shows a systematic increase with the same. For $L^2 = 1.01$, the drag correction coefficient from FENE-P also remains close to unity and is thus consistent with the equivalent CY model. To explore this further, we study the velocity and stress contours for $L^2$=1.1 and 1.01 in Fig. 10. Figs. 10 (a) - (c) and (d) – (f) show results for $L^2 = 1.1$ and 1.01, respectively. The stress contours, especially for the polymeric contribution, depict much higher degrees of asymmetry for $L^2 = 1.1$, relative to $L^2 = 1.01$. Obviously, this has to do with the reduction in the elastic nature for $L^2 = 1.01$. Note that a smaller value of $L^2$ means that the spring connecting the beads is not quite extensible. This behaviour also manifests itself as reduced FNSD (with lower $L^2$) (shown in the inset of Fig. 6). This effect is highly pronounced for $L^2$=1.01. Note that, as the extensibility parameter reduces for the FENE-P, it should get closer to the CY-GNF Model. In the limit $L^2 \to 1$, they will be the same. Consequently, as $L^2$ approaches unity, the velocity, total, and polymeric stress contours contour must all become symmetric about the sphere. This symmetry can be observed in the right column of Fig. 10, as mentioned before. Hence, only the limit in which the polymer chain of the fluid being studied is nearly inextensible, is it justified to employ the equivalent GNF model for fluid flow analysis. Note, however, in all practical applications, the polymer chain will be highly extensible, with a sufficiently high $L^2$.

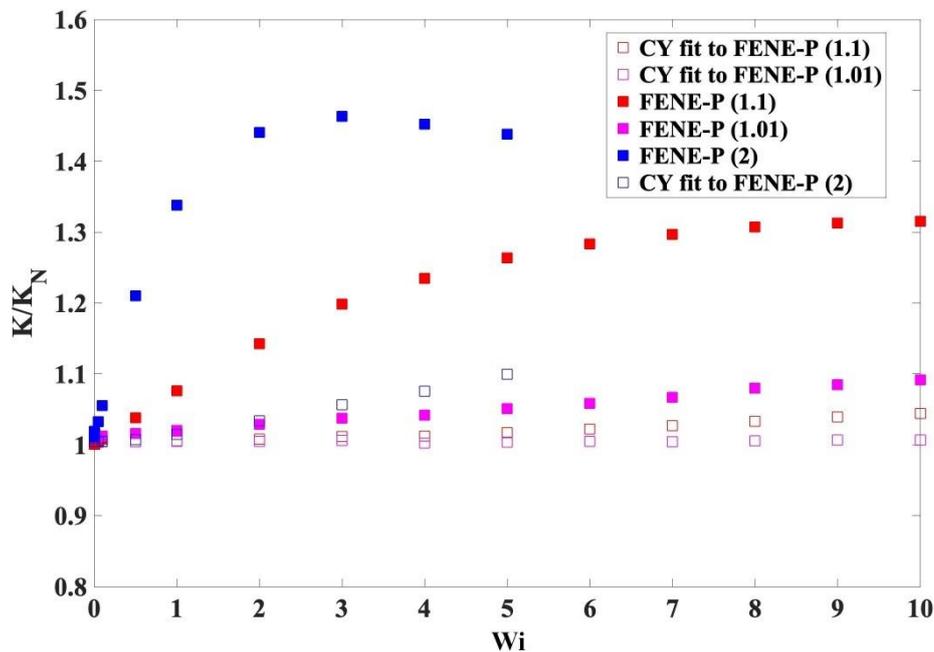

**FIGURE 9** – Variation of the drag correction coefficient with Wi for different FENE-P fluids and their equivalent CY-GNF models.

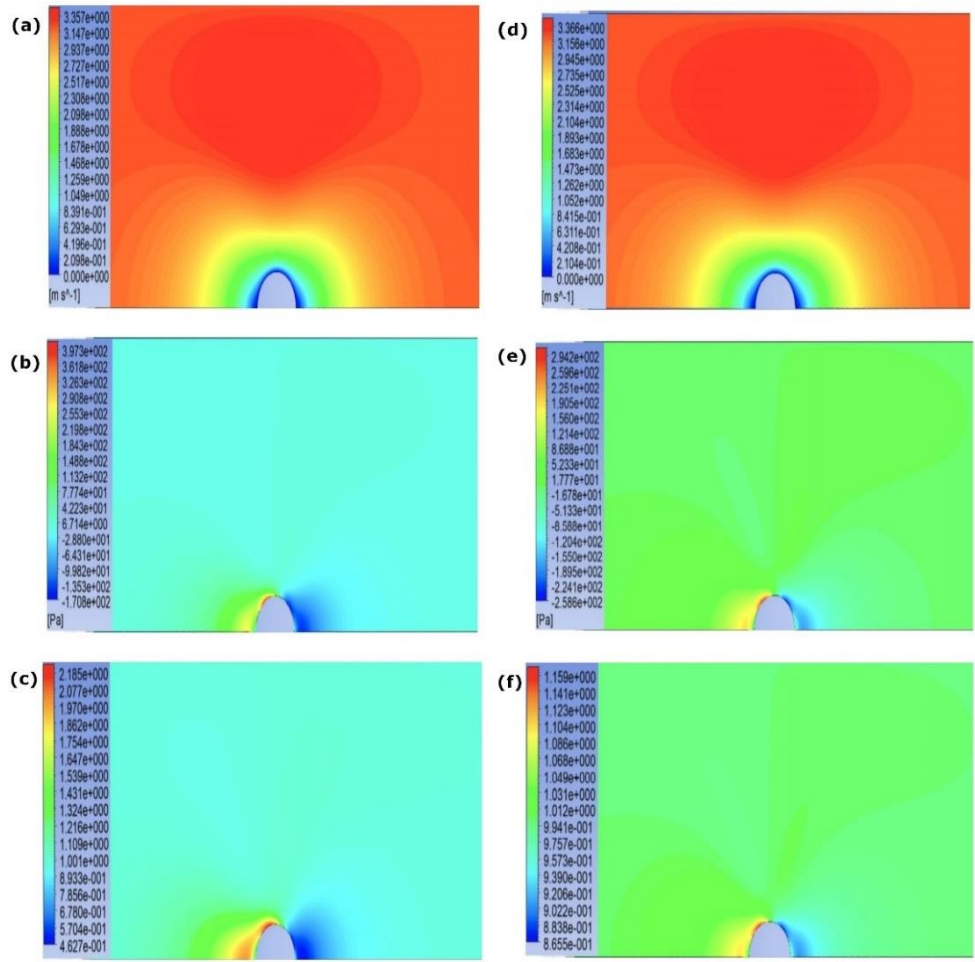

**FIGURE 10** – (a), (b), and (c) represent the velocity contour, zz-component of total stress contour and zz-component of polymeric stress tensor for the FENE-P model with $L^2 = 1.1$. (d), (e), and (f) show the same but for $L^2 = 1.01$.

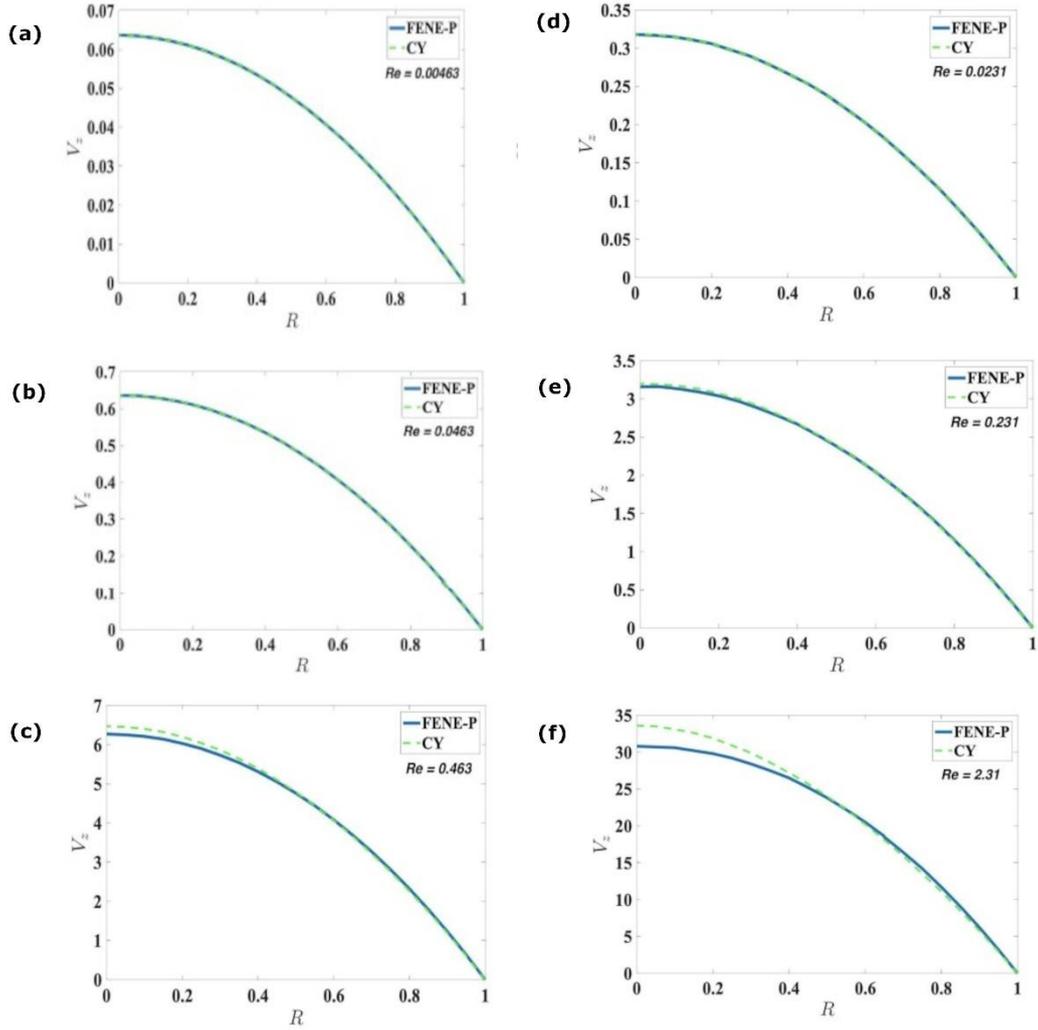

**FIGURE 11**: Velocity profiles for flow in a pipe for the FENE-P fluid ($L^2 = 1.1$) and the equivalent CY-GNF model for (a) Re = 0.00463, (b) Re = 0.0463, (c) Re = 0.463, (d) Re = 0.0231, (e) Re = 0.231 and (f) Re = 2.31. Here R represents the radial coordinate.

Hence, the inability of the GNF models to capture the elastic nature of the fluid is the primary source of the disagreements. For external flows, the velocity field typically has extensional (i.e. stretching) components. Thus, the elastic nature of the fluid becomes important. However, if any flow is primarily shearing in nature, with minimal extensional components, any equivalent model that captures the shear-thinning effect should provide satisfactory predictions. To check this, we simulated the flow in a pipe for both FENE-P and the equivalent CY-GNF. The velocity profiles are shown in Fig. 11. As expected, the profiles agree well for the entire cross-section, within the range of flow rates studied. However, the FENE-P model shows a lower maximum velocity near the centerline of the pipe, as flow rates are increased.

### 3.2 Comparison of FENE-P and GNF-X model

From the previous section, we observed that the GNF fails to replicate the behavior of the FENE-P model due to its inability to capture the elastic effects. For any general flow field, especially around an object, these effects become significant as the polymer chains get stretched and compressed while flowing past the body. This has been shown experimentally for the flow of DNA chains around a cylinder [19]. One possible solution is to identify the extensional and shearing parts locally in a flow field and estimate the stress accordingly. Recently, Tseng et al. [14] formulated a hybrid viscosity model, called the GNF-X, which linearly combines the shear and elongational viscosity to obtain a mixed viscosity, by estimating the shear and extensional strains, respectively. They highlighted its success for a 2D contraction flow by predicting the entry vortex effect for polymer melts. However, this has not been extensively tested, particularly for the class of external flows like the flow around a sphere, which is considered in this study. However, the premise of the GNF-X is promising, since it attempts to incorporate the extensional effects, which were lacking for the GNF family of models. If successful, it might prove to be a substitute to more complicated constitutive models like FENE-P. In what follows, the GNF-X model is tested for flow around a sphere, relative to the FENE-P.

The GNF-X model estimates the local viscosity using the Eqns. (6) and (7), as mentioned in section 2.1.2. The shear ($\eta_S$) and extensional viscosity ($\eta_E$) are computed using the values of $\dot{\gamma}'_S$ and $\dot{\gamma}'_E$, respectively. *W* represents the weight function to estimate the local viscosity, using the extensional and shear parts. This splitting of the total strain rate into extension and shear rates is performed through the transformation of the rate-of-strain tensor to the streamline coordinate system. This transformation yields the principal strain tensor, which can be easily decomposed into the principal shear and elongation tensor. The transformation, or the rotation matrix, is given as:

$$\vec{R} = \begin{bmatrix} \vec{t} \\ \vec{n} \\ \vec{b} \end{bmatrix} \quad (21)$$

Here $\vec{t}$, $\vec{n}$, and $\vec{b}$ are the unit vector along the tangent, normal, and the binormal direction of the curve, respectively. In the streamline coordinate system, the tangential direction is given by the direction of the velocity vector (by the definition of the streamline). The normal vector at any point on the curve is pointing outward, away from the curvature. The binormal vector is defined as the cross-product of $\vec{t}$ and $\vec{n}$.

The rotation is obtained as follows

$$\vec{D}' = \vec{R}^T \cdot \vec{D} \cdot \vec{R} \quad (22)$$

Where the rotation matrix is defined in Eqn. (21). After rotation, the principal rate-of-strain tensor $\vec{D}'$ can be split into a principal shear and a principal extension tensor as follows:

$$\vec{D}' = \vec{\dot{\gamma}}'_S + \vec{\dot{\gamma}}'_E \quad (23)$$

If

$$\vec{D}' = \begin{bmatrix} d'_{11} & d'_{12} & d'_{13} \\ d'_{21} & d'_{22} & d'_{23} \\ d'_{31} & d'_{32} & d'_{33} \end{bmatrix} \quad (24)$$

then we have the following for the shear and extensional strain rates:

$$\vec{\dot{\gamma}}'_S = \begin{bmatrix} 0 & d'_{12} & d'_{13} \\ d'_{21} & 0 & d'_{23} \\ d'_{31} & d'_{32} & 0 \end{bmatrix} \quad (25)$$

$$\overrightarrow{\dot\gamma'_E} = \begin{bmatrix} d'_{11} & 0 & 0 \\ 0 & d'_{22} & 0 \\ 0 & 0 & d'_{33} \end{bmatrix} \tag{26}$$

Then, we obtain the principal deformation rates using the following equations:

$$\dot\gamma' = \sqrt{2\overrightarrow{\mathbf{D}'}:\overrightarrow{\mathbf{D}'}} \tag{27}$$

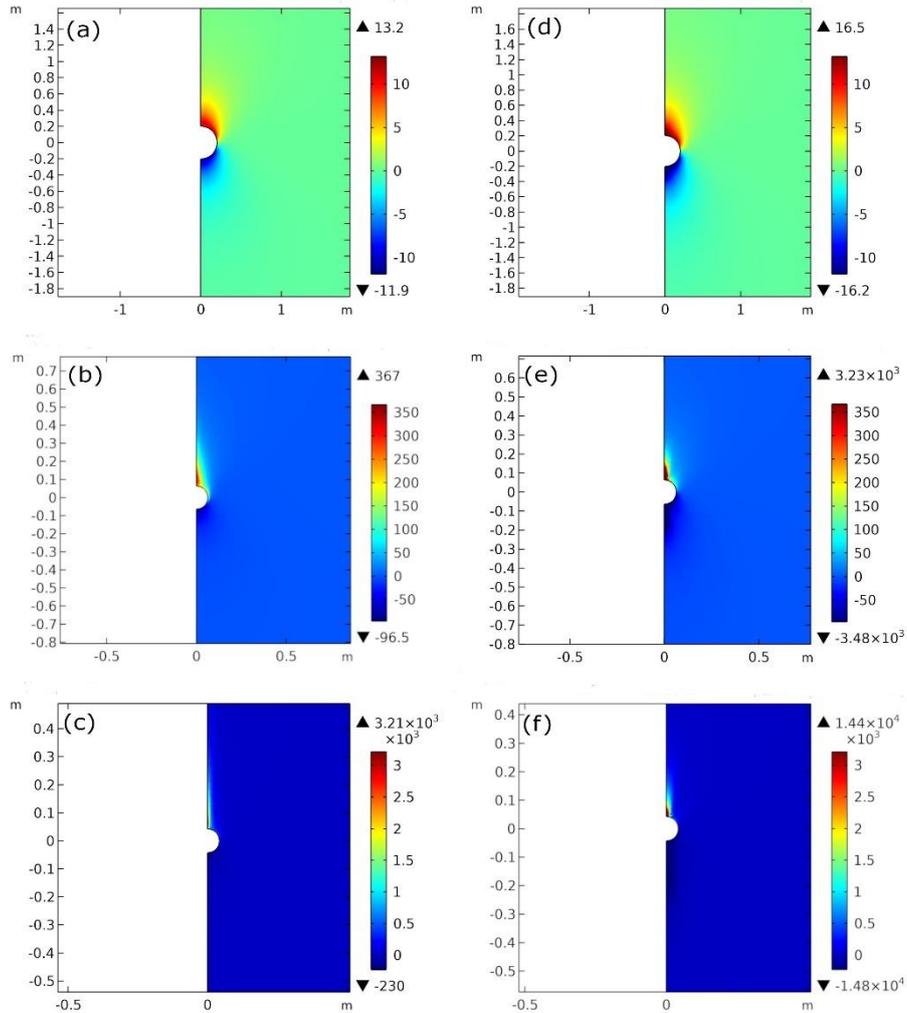

**FIGURE 12:** (a), (b) and (c) represents the contour of zz-component of total stress tensor for the FENE-P model at Re = 0.001 and Wi of 0.1, 1 and 2.24 respectively. (d), (e) and (f) represent the same case as (a), (b), (c) but for the GNF-X model. Parameters for the FENE-P model are mentioned in Table 4.

$$\dot\gamma'_S = \sqrt{2\overrightarrow{\dot\gamma'_S}:\overrightarrow{\dot\gamma'_S}} \tag{28}$$

$$\dot\gamma'_E = \sqrt{2\overrightarrow{\dot\gamma'_E}:\overrightarrow{\dot\gamma'_E}} \tag{29}$$

The strain rate in the original coordinate system is equal to the principal strain rate, as noted in the original study of GNF-X [14]

$$\dot\gamma = \dot\gamma' \tag{30}$$

The relation of the principal strain rate with the deformation rates is given as:

$$\dot{\gamma}'^2 = \dot{\gamma}'^2_E + \dot{\gamma}'^2_S \tag{31}$$

which can be easily observed from the way the quantities are defined above. Thus, starting with an arbitrary local strain rate, we have obtained the extensional and shear strains, which can be further used to compute the extensional and shear viscosities and the weight function.

In this study, this technique is applied for flow around a sphere, using a 2D-axisymmetric geometry. Note, the definition of the normal vector at any point, in 3D space, is not unique for the GNF-X model. However, for the axisymmetric 2D setup, this is uniquely defined and is directed along the radial direction at any point. For the calculation of the extensional viscosity, we used the steady uniaxial extensional flow setup of a FENE-P fluid, as described in [20]. A mathematical expression is fitted to the normalized extensional viscosity since it is a function of the extension rate and the extensibility parameter ($L^2$). The mathematical expression for fitting the normalized viscosity data has been inspired by the formulation of Sarkar and Gupta [21].

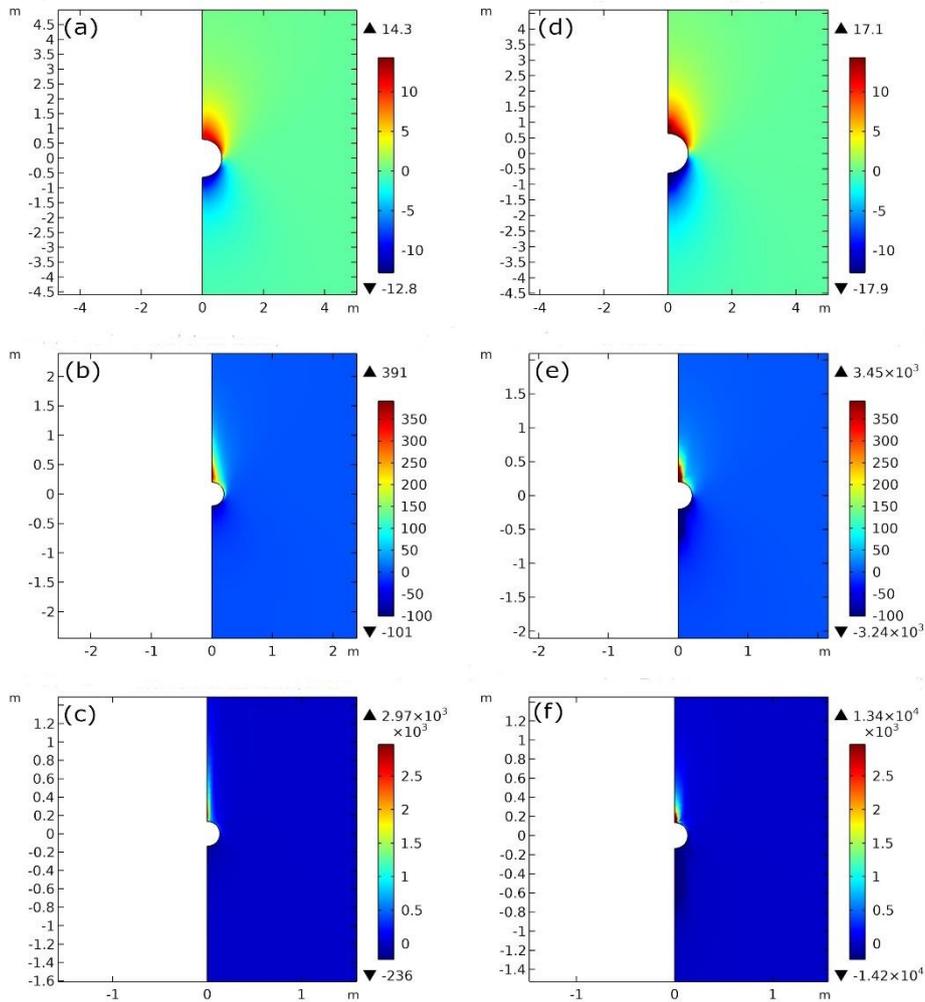

**FIGURE 13:** (a), (b) and (c) represents the contour of zz-component of total stress tensor for the FENE-P model at Re = 0.01 and Wi of 0.1, 1 and 2.24 respectively. (d), (e) and (f) represent the same as (a), (b), (c), but for the equivalent GNF-X model. The parameters for the FENE-P model are mentioned in Table 4.

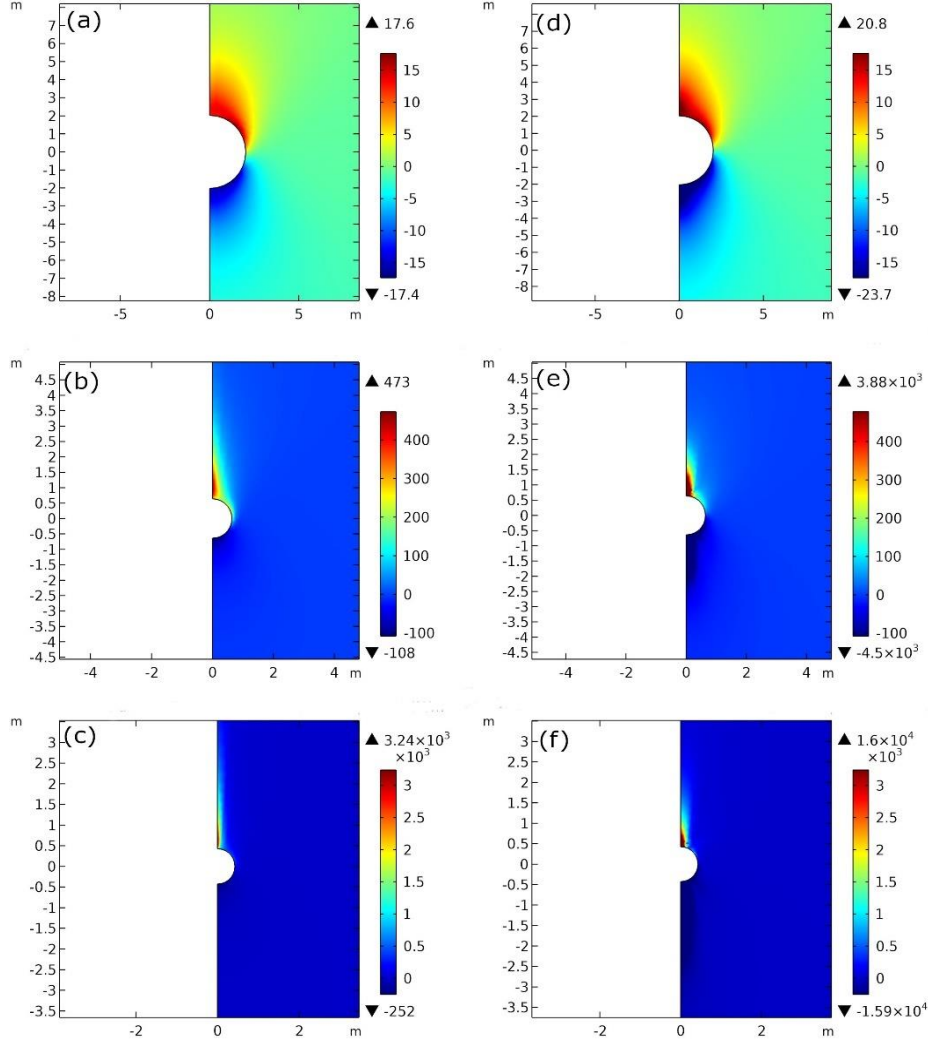

**FIGURE 14:** (a), (b) and (c) represents the contour of the zz-component of total stress tensor for the FENE-P model at Re = 0.1 and Wi of 0.1, 1 and 2.24 respectively. (d), (e) and (f) represent the same as (a), (b), (c), but for the equivalent GNF-X model. Parameters for the FENE-P model are mentioned in Table 4.

The expression used for fitting the normalized extensional viscosity is given as:

$$\eta_{norm} = 1 + \frac{a}{(1+(b\dot{\varepsilon})^{-2})^c} \tag{32}$$

Where the normalized extensional viscosity $\eta_{norm}$ is defined as:

$$\eta_{norm} = \frac{\eta - 3\eta_S}{3\eta_0 - 3\eta_S} \tag{33}$$

Here, $a$, $b$ and $c$ are the parameters which are obtained by fitting this curve to the data of extensional viscosity obtained from simulations, $\eta_S$ is the viscosity of the solvent and $\eta_0$ is the zero-shear viscosity of the polymer solution. The details of fitting procedure is given elsewhere [22].

To understand its performance, we compare the stress profiles obtained using this model and FENE-P. For a detailed comparison, we have taken three levels of Reynolds Number (Re), which are 0.001, 0.01 and 0.1. For each Re, we have taken three levels of Weissenberg Number (Wi), which are 0.1, 1 and 2.24. These three values of Wi denote the different regimes of stretching of the polymer chains, with Wi = 0.1 expected to show negligible stretching, Wi = 1 represents moderate stretching and Wi = 2.24 is strong enough to significantly stretch the chains. The values of the parameters used for the FENE-P model in these simulations are mentioned in Table 4.

Table 4: Parameters used for FENE-P in the simulations for the comparison between GNF-X and FENE-P model

| | |
|---|---|
| $\rho$ (kg/m$^3$) | 1 |
| $\eta_p$ (Pa.s) | 13.76 |
| $\eta_s$ (Pa.s) | 8.12 |
| $\lambda$ (s) | 0.749 |
| $L^2$ (dimensionless) | 500 |

Figure 12 compares the stress profiles of GNF-X and FENE-P For Re = 0.001, with Wi being 0.1 (a and d), 1 (b and e) and 2.24 (c and f). Parts (a) – (c) and (d) – (f) show the results for the FENE-P and GNF-X, respectively. The stress profiles remain similar in low Wi but differ drastically as Wi is increased, when considerable chain stretching occurs. For low Wi, the profiles are similar, with some high-stress zones occurring with the GNF-X model. At higher Wi, the profiles are quite different, with the presence of high-stress zones with the GNF-X model near the stagnation point in the rear as well as the leading end of the sphere. The stress magnitudes at these stagnation points are nearly five times higher relative to FENE-P. Also, the GNF-X model, in its native form, clearly fails to show any asymmetry in the stress contours between the leading and rear sides of the sphere.

The analysis is continued further for higher Re values of 0.01 (Fig. 13) and 0.1 (Fig. 14). Overall, the trends are similar to those observed in Fig. 12. There exist significant differences in the stress profiles at moderate-to-high Wi, for all values of Re. High-stress zones exist at all flowrates and there is an increased extension at the rear stagnation point for both the models, thereby generating a higher stress. However, this does not explain the high stress observed in the leading stagnation point for GNF-X, which is not observed in FENE-P since the polymer chain stretch is expected to be low in this region. Thus, the current formulation of GNF-X is incapable of replacing the FENE-P, since there exists major differences in the predictions for this relatively simple problem, even qualitatively. However, it does account for the extensional component of the flow to some extent, thereby showing some promise towards predicting the asymmetry in the profiles (as observed with the FENE-P model). In what follows, we formulate a modified version of the GNF-X model that is able to deal effectively with such inconsistencies and shows a great promise towards approximating FENE-P.

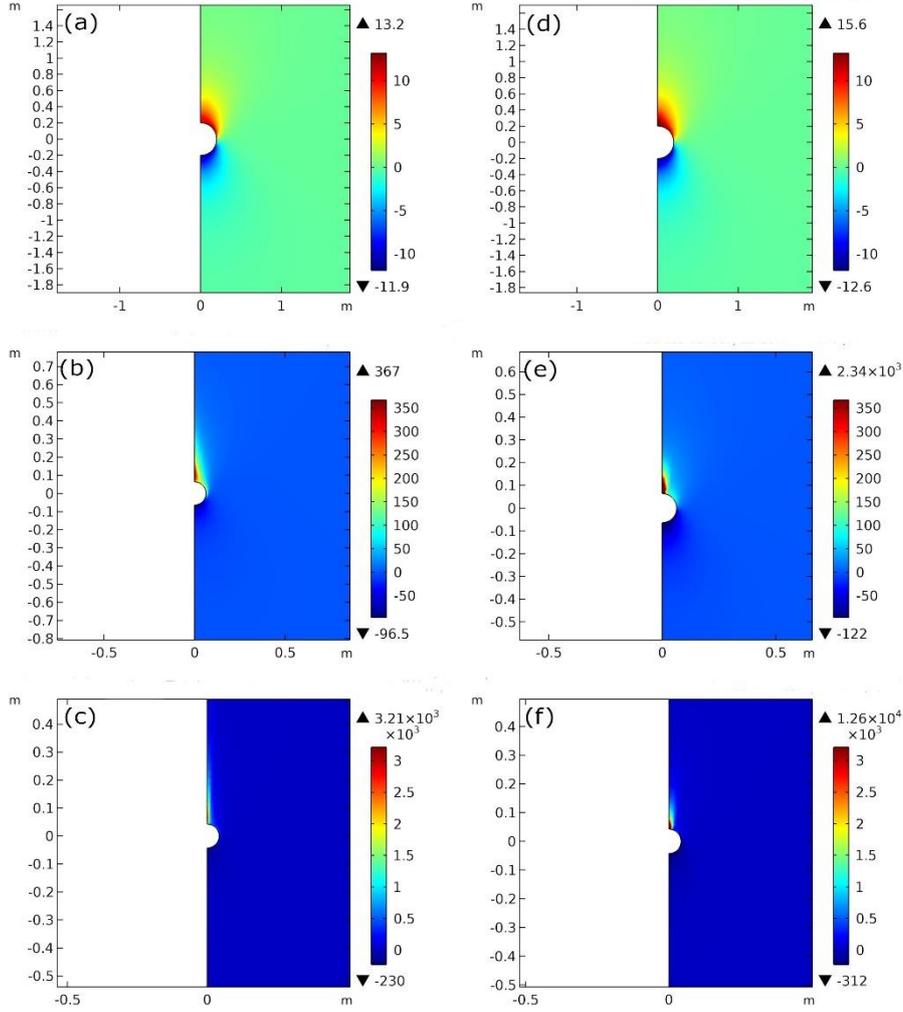

**FIGURE 15:** (a), (b) and (c) represents the contour of the zz-component of total stress tensor for the FENE-P model at Re = 0.001 and Wi of 0.1, 1 and 2.24 respectively. (d), (e) and (f) represent the same as (a), (b), (c), but for the equivalent GNF-XM model. The parameters for the FENE-P model are mentioned in Table 4.

3.3 Comparison between FENE-P and GNF-XM

As discussed in the previous section, the GNF-X model shows larger stresses in the region of the leading stagnation point for moderate to high Wi, in sharp contrast to the FENE-P model. Additionally, the stress values in the leading and rear stagnation regions are similar. Thus, it fails to exhibit the asymmetry of the stress values in the leading and rear regions of the sphere. Qualitatively, the primary reason behind this trend is the failure to differentiate between the nature of stretching that the polymer chains experience at leading and rear regions. As discussed earlier, the chains are compressed in the leading region of the sphere and stretched in the rear region. However, the mathematical formulation of the GNF-X is unable to capture this aspect. We thus propose a novel modification to the GNF-X model in order to differentiate between local compression and extension while calculating the extension rate in the streamline coordinate system. Here, we use the notion that the spring constant increases at large chain stretch but remains Hookean even at large compression (relative to the equilibrium configuration). Thus, the extensional viscosity for large compressive strain rates is much different (and equal to that at equilibrium) from that for similarly large extensional strain rates (much larger than the value at equilibrium). Thus, mathematically, we put a constraint that allows the extensional viscosity to be affected only due to chain elongation and not compression. This also corroborates well with the observations of Chilcott and Rallison [23] much earlier. They proposed, based on their experiments, that the main polymeric contributions lie near the rear stagnation point and its further downstream region. This is similar to our observations with the FENE-P, which predicts higher stresses in the downstream region. Clearly, the original formulation of extensional viscosity has an issue, which can potentially affect the predictions for any external flow problem. Also, note that the definition of the extensional viscosity used earlier is correct from a rheological point of view, with the expression being for uniaxial extension. However, this same definition, when applied for a Newtonian fluid, would predict an extensional viscosity that is three times of the shear viscosity. However, since the fluid is Newtonian, the viscosity value would always be constant and can't change with the type of flow. To remove this inconsistency, we also propose that the extension viscosity used throughout in this study must be scaled down by factor of 3.

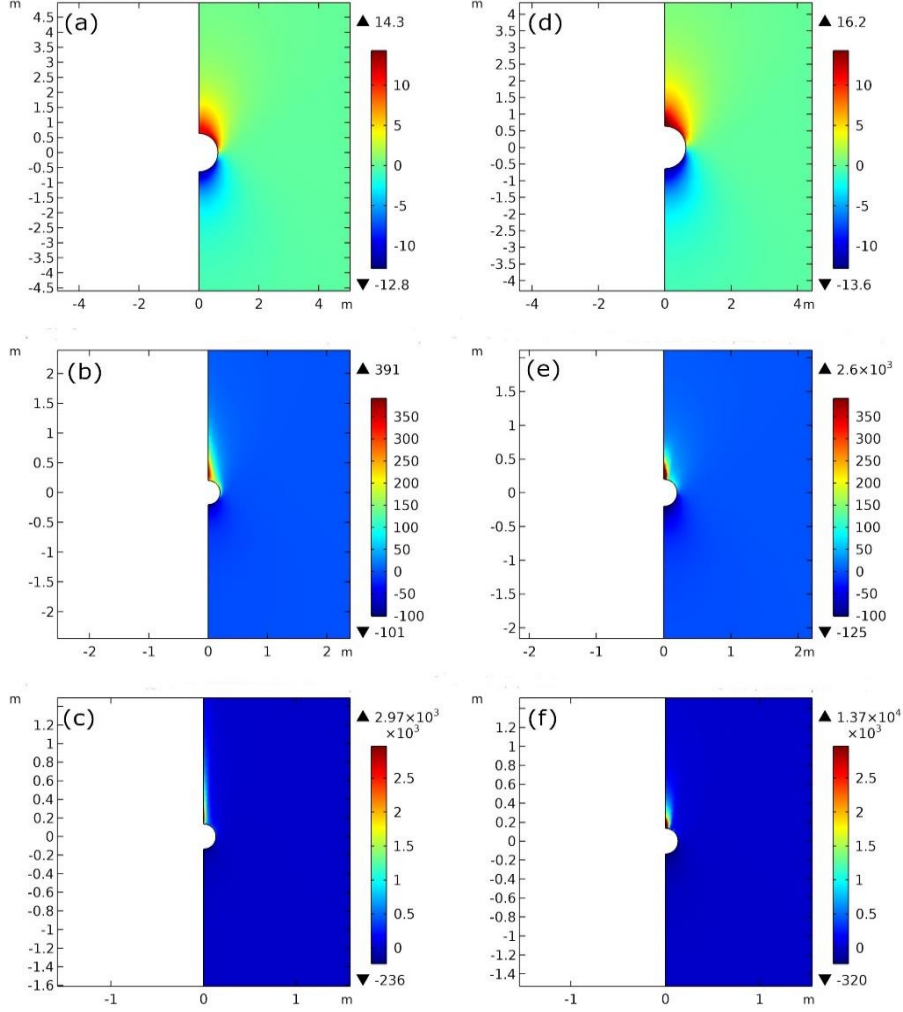

**FIGURE 16:** (a), (b) and (c) represents the contour of the zz-component of total stress tensor for FENE-P model at Re = 0.01 and Wi of 0.1, 1 and 2.24 respectively. (d), (e) and (f) represent the same as (a), (b), (c), but for the equivalent GNF-XM model. The parameters for the FENE-P model are mentioned in Table 4.

To summarize, the calculation of the extensional strain for the modified GNF-X model is described mathematically as:

$$\dot{\gamma}'_E = d'_{11} \quad \text{if} \quad d'_{11} > 0 \tag{34}$$

$$\dot{\gamma}'_E = 0 \quad \quad \text{if} \quad d'_{11} < 0$$

Here $d'_{11}$ represents the extensional strain along the streamline direction. A positive value of this quantity represents extension while negative represents compression. This definition is expected to make the model exhibit elastic characteristics only in the rear region, where there is considerable extension of the polymer chains. When the flow is compressional, as in the leading region, the effect on the viscosity is assumed to be small enough so that it can be neglected for future calculations. Thus, in the leading region, the Newtonian solvent viscosity and the shear component of the viscosity due to the polymer dominates. Henceforth, we call this model as GNF-XM (GNF-X modified). In what follows, we show how such simple modifications lead to significantly better predictions.

To compare the GNF-XM to FENE-P, we compare their stress profiles. For this, three levels of Re (0.001, 0.01 and 0.1), with three Wi (0.1, 1 and 2.24) are considered, as before. The parameters for the FENE-P model are mentioned in Table 4. The normal stress profiles are shown for Re = 0.001, 0.01 and 0.1 in Figures 15, 16 and 17.

Figure 15 compares the stress profiles of GNF-XM and FENE-P For Re = 0.001, with Wi being 0.1 (a and d), 1 (b and e) and 2.24 (c and f). Parts (a) – (c) and (d) – (f) show the results for the FENE-P and

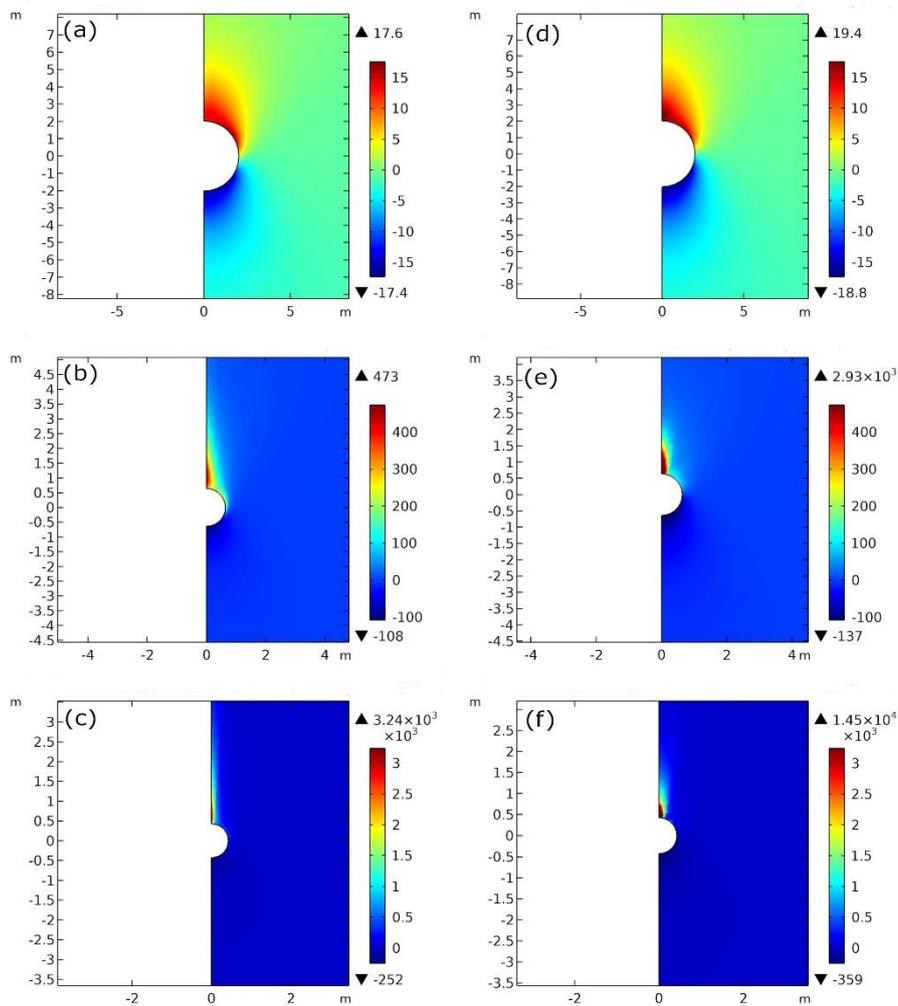

**FIGURE 17:** (a), (b) and (c) represents the contour of the zz-component of total stress tensor for the FENE-P model at Re = 0.1 and Wi of 0.1, 1 and 2.24 respectively. (d), (e) and (f) represent the same as (a), (b), (c), but for the equivalent GNF-XM model. The parameters for the FENE-P model are mentioned in Table 4.

GNF-XM, respectively. We observe that at low Wi, the profiles are similar as there is no stretching. At moderate and high Wi, the GNF-XM model agrees well with the trends of FENE-P, with lower stress in the leading region and higher stress in the rear region. This highlights the success of this model in differentiating between stretching and compression and its effect on viscosity. It clearly shows the asymmtery in stress profiles at moderate and high Wi, consistent with FENE-P trends.

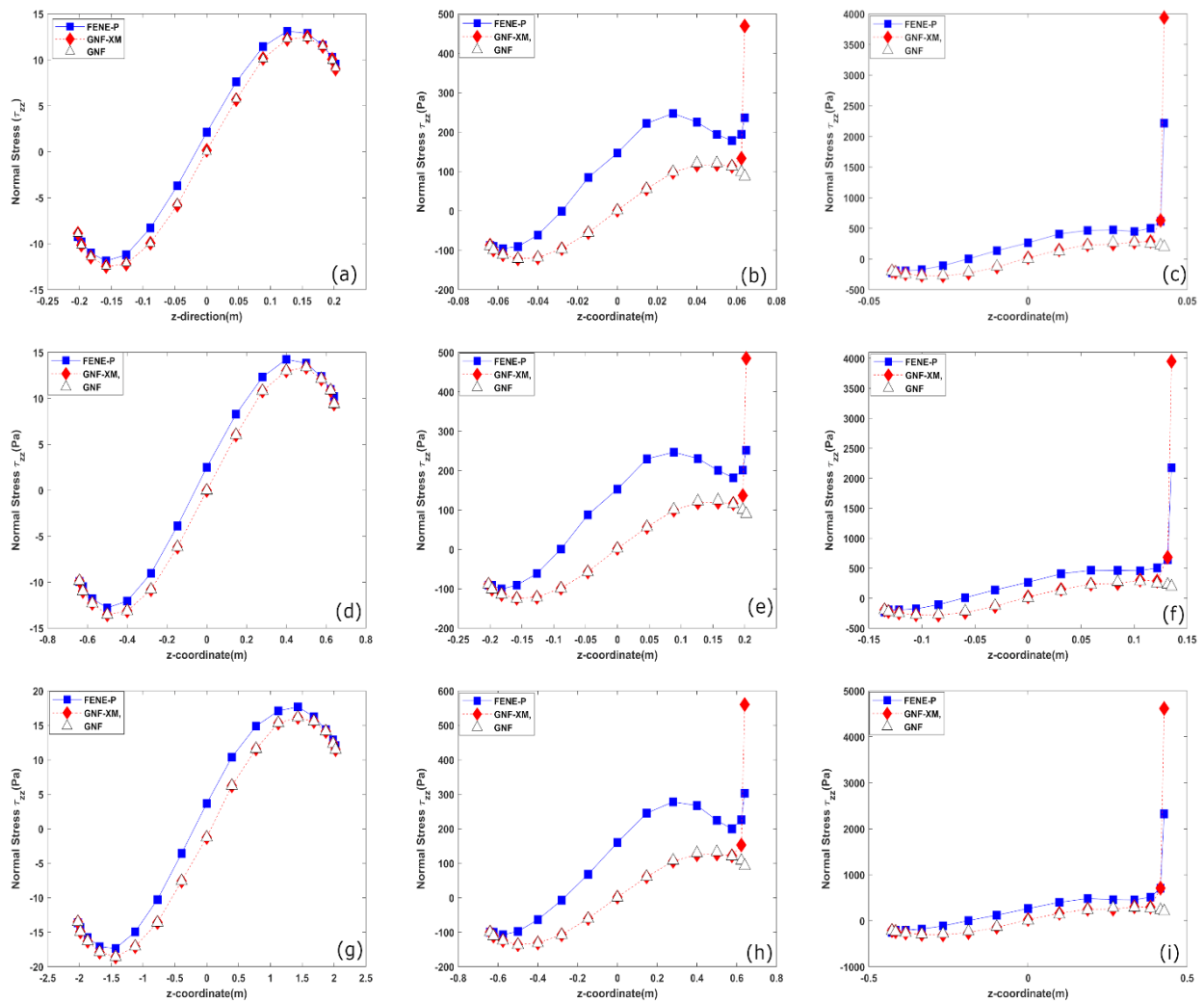

**FIGURE 18:** The variation of the normal stress on the surface of a sphere for the FENE-P and the equivalent GNF and GNF-XM models, with the varying Wi and Re. (a)-(c) shows the variation of stress with Wi for Re = 0.001. (d)-(f) shows the same for Re = 0.01, and (g)-(i) shows the variation for Re = 0.1. Thus, from top to bottom, the Re values are 0.001, 0.01 and 0.1. From left to right, the Wi values are 0.1, 1 and 2.24. The lines are a guide to the eye.

Further comparisons of stress profiles is performed for Re values of 0.01 and 0.1 (Figs. 16 and 17). We observe similar trends as for Re = 0.001, with an increase of the stress magnitudes. Thus, overall, the GNF-XM is largely successful predicting the asymmetry in stress profiles. The stress magnitudes agree well with FENE-P for all ranges of Wi and Re at all locations except at the rear stagnation point, where the GNF-XM predictions are higher than FENE-P. For a deeper analysis of the success of the modified model to predict stresses, relative to GNF-X and GNF models, keeping FENE-P as the benchmark, we study the variation of stress along the surface of the sphere for all the cases of Re (0.001, 0.01 and 0.1) and Wi (0.1, 1 and 2.24). This is important for the following reasons:

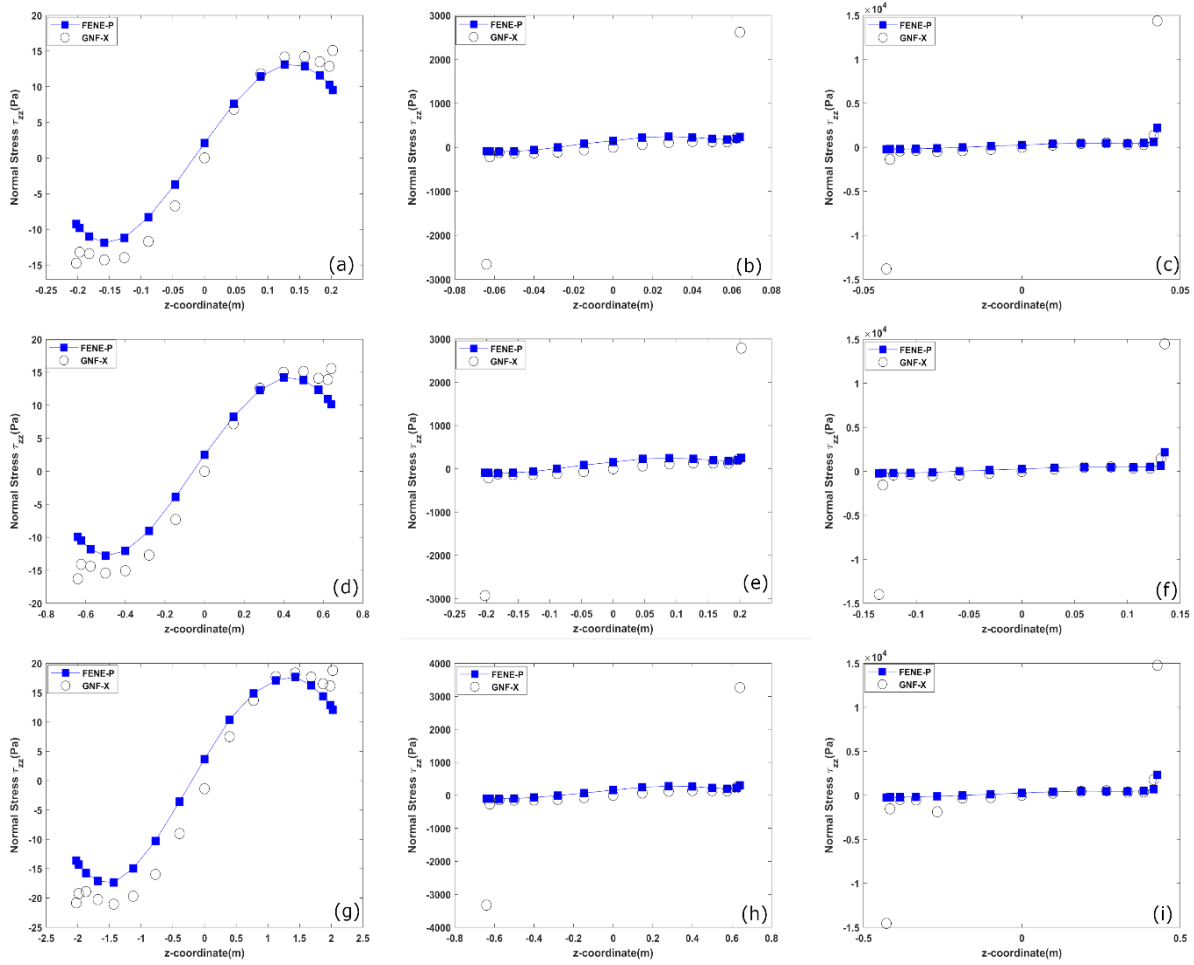

**FIGURE 19**: The variation of the normal stress on the surface of a sphere for the FENE-P and the equivalent GNF-X, with a variation of Wi and Re. (a)-(c) shows the variation of stress with Wi for Re = 0.001. (d)-(f) shows the same for Re = 0.01, and (g)-(i) shows the variation for Re = 0.1. Thus, from top to bottom, the Re values are 0.001, 0.01 and 0.1. From left to right, the Wi values are 0.1, 1 and 2.24. The lines are a guide to the eye.

1. The contour plots provide an overall variation of stress in the entire domain and are less informative for the same on the surface of the sphere.
2. The variation of stress on the surface is extremely important for engineering applications owing to its direct relevance to the drag force and the drag coefficient.

For completeness, we also analyzed the variation of the drag correction coefficient ($\chi$) with Wi for all the models considered in this study. Note, the drag correction coefficient is defined in Eqn. 14. Such measurements have been traditionally used since 1992, after Walters and Tanner [24] proposed a generalized trend of drag correction coefficient vs Wi for different classes of shear thinning polymer solutions and Boger fluids for flow around a sphere in the absence of wall effects. Since that study, many researchers have confirmed this trend experimentally and used it as a benchmark for experimental studies and constitutive equations. For this study, the same analysis for different models is shown in Figure 20. The simulations for obtaining the drag correction coefficient for the FENE-P model are performed in OpenFOAM V7.0 with an add-on tool named rheoTool V4.0[25]. A similarly fine mesh is taken in OpenFOAM as our simulations in COMSOL. These drag coefficient computations are performed in OpenFOAM to be consistent with similar calculations in a recent study [26], that uses OpenFOAM instead of COMSOL.

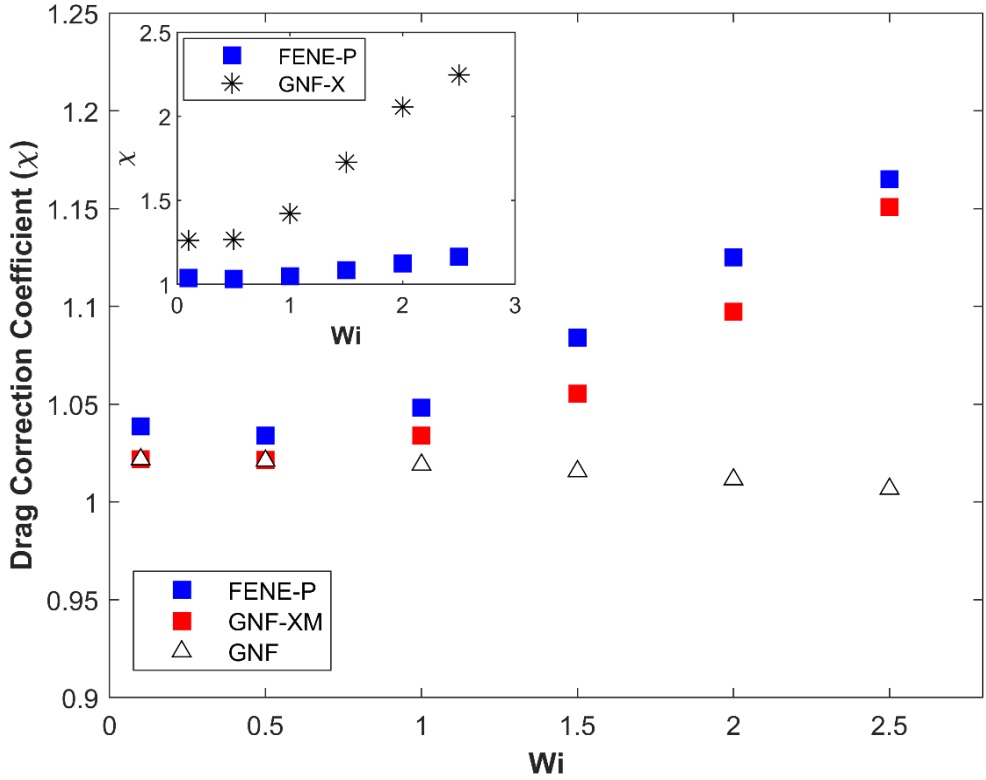

**FIGURE 20:** The variation of the drag correction coefficient ($\chi$) with Weissenberg Number (Wi) for FENE-P and the equivalent GNF-XM, and GNF models for an unconfined flow around a sphere. The inset shows the variation of $\chi$ with Wi for the FENE-P and the equivalent GNF-X model. Notice that the GNF-XM is the only model that provides similar predictions as FENE-P for all Wi considered here.

Figures (18) and (19) show extremely high stresses predicted by GNF-X at both the front and rear stagnation points. However, at all other points, similar stress values are observed for all models i.e., the equivalent GNF, GNF-X and GNF-XM. The equivalent GNF fails to show the high stress at the rear stagnation point due to its inability to capture chain stretching, as discussed before. At low Wi, when there is negligible stretching, all models agree well, with the exception being GNF-X, which shows higher values of normal stress at both the stagnation points relative to the others. At moderate Wi, GNF-X continues to show high stress magnitudes at both the leading and rear stagnation points. The underlying reason has been discussed earlier, which provided the motivation for proposing the modified GNF-XM model. Quite interestingly, GNF-XM is able to correct these issues and the trends agree well with FENE-P. However, the stress magnitude at the rear stagnation point is larger than FENE-P (Fig. 18), although in much better agreement than the same from GNF-X (Fig. 19).

On closer inspection, in the regions of low extension rates, i.e., in the region between the stagnation points, the GNF, GNF-X and GNF-XM models agree with each other across all the values of Wi and Re. Here, note that the GNF-X, as well as the GNF-XM model, essentially adds a term that accounts for an increase in viscosity due to the stretching of polymer chains. Thus, in the regions where the extensional rates are lower, the shear viscosity contribution will dominate over the extensional viscosity, effectively yielding the same equivalent GNF model. At moderate stretching, i.e., at Wi = 1, the stresses predicted by the FENE-P and the GNF class of models are somewhat different in the regions of low extension. The FENE-P model predicts higher stress in these regions, relative to all GNF models. This trend is also visible at higher Wi. This analysis shows that the GNF, GNF-X and GNF-XM behave similarly due to their inherent equivalence in the shear viscosity, and differences arise as the flow field becomes more extensional in nature. This is primarily the case near the leading and rear stagnation points.

Next, we analyze the predicted drag correction coefficients for all the models used here. This is shown an a function of Wi in Figure 20. For the equivalent GNF model, the drag correction coefficient remains about unity with a slight dip at higher Wi. This is expected since GNF models do not incorporate any extensional contribution and hence are incapable to alter the drag even when the flow has extensional components. The slight dip is due to the shear thinning effect of the GNF model, which causes the viscosity to drop with an increase in shear rate, resulting in a decrease of the drag force. For the FENE-P model, the drag correction coefficient dips slightly and then increases at larger Wi. This can be attributed to the dominance of elastic stresses at higher Wi, which increases the drag force and the drag correction coefficient with respect to the Newtonian fluid. Quite significantly, the GNF-XM model predictions agree quite well with those for FENE-P, with slightly lower values. This is a massive improvement over the GNF-X model, whose predictions are much larger than FENE-P (inset of Fig. 20). Thus, for this property of great practical relevance, the GNF-XM, formulated with simple modifications, performs much better than both an equivalent GNF and the original GNF-X.

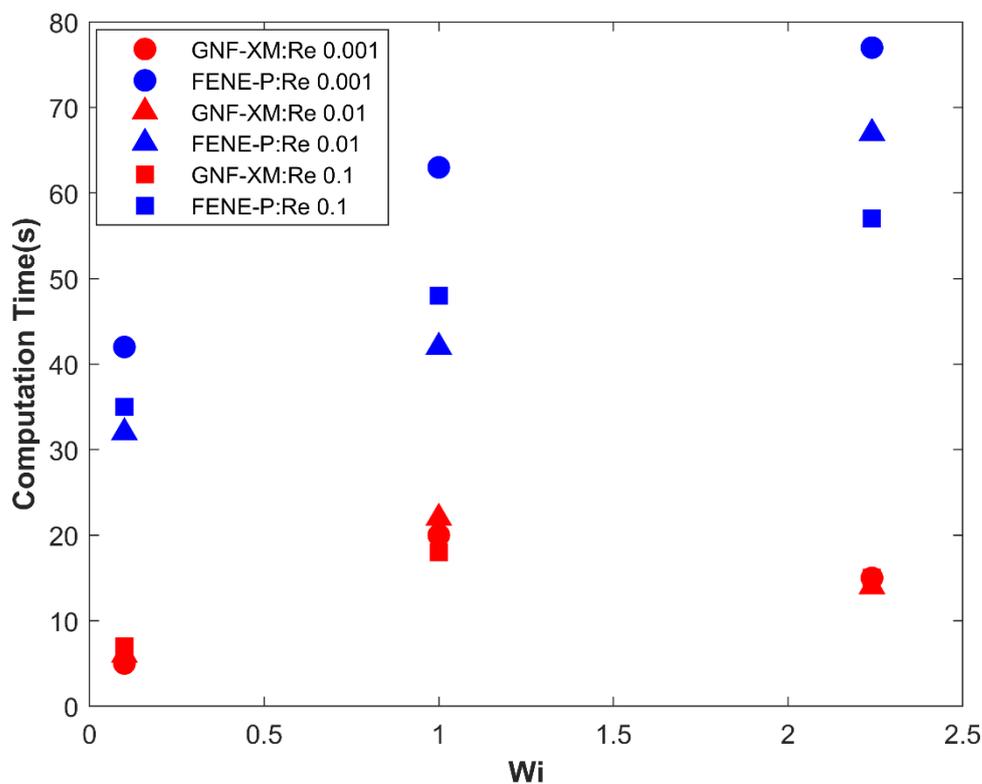

**FIGURE 21:** The variation of the computational times required to solve the different cases of Re and Wi, for the FENE-P and the equivalent GNF-XM model, on a Dell laptop (Intel i5 8$^{th}$ generation processor, with 8 GB RAM).

Further, having compared the accuracy for this flow, we analyze the speed for solving the flow field. For this, the computational time taken to solve such cases in COMSOL, for all the cases of Re and Wi using the FENE-P and the GNF-XM, are measured. This is shown in Figure 21. For all simulations shown here, a Dell laptop with an Intel i5 8$^{th}$ generation processor and 8 GB RAM is used. Clearly, the GNF-XM model takes much lower time to compute any of the simulation cases used in this study, with savings of 2-8 times (almost an order of magnitude for higher Wi). Further, the FENE-P computational time increases with Wi (presumably the solver takes longer to converge with higher degrees of chain stretching), whereas the GNF-XM roughly saturates to a plateau. Thus, the GNF-XM agrees well with the FENE-P (much better agreement than the equivalent GNF models) and yet takes much lower computational time. The key takeaway message, thus, is that the GNF-XM can be positioned as a cheaper alternative to the FENE-P for various applications, especially when geometries are complicated, and flowrates are high enough to stretch the chains. Such complex geometries are normally found in industrial applications.

Thus, our results clearly indicate that the GNF-XM model represents a hugely promising step towards obtaining an appropriate cheaper alternative of highly non-linear molecular models like the FENE-P. Note, however, it still has room for improvement. This pertains to the lower stresses shown by GNF-XM particularly in the intermediate regions of the sphere, where the extensional viscosity component is less dominant. Note, one important advantage of the GNF-X (or GNF-XM) formalism is its universality in terms of application. That is, any given constitutive equation, however complicated to solve numerically, can be reduced to its equivalent GNF-X model. This can be achieved by following the steps in this manuscript. First, the constitutive model can be used to generate the shear and extensional viscosity data. Next, these can be fitted to a GNF model for the shear component and a mathematical expression can be fitted to the extensional viscosity data. The two together, with a suitable weight function, gives the resulting GNF-X model.

## 4. Conclusion

To summarize, in this study, we highlight the possibility of a computationally simpler alternative to the FENE-P. The FENE-P model developed decades ago by Bird and coworkers[9] is rich in physics but difficult to converge numerically for various flows involving polymer solutions. The study primarily builds upon the GNF framework, encompassing simple shear thinning models and the GNF-X formalism, where the viscosity is tuned by using the local extensional and shear strain rates. Our study clearly indicates the following:

- The predictions from an equivalent GNF model (CY in this study) is not satisfactory for external flows (i.e. flow around an object, which has extensional components). However, the results are nearly identical for an internal flow (i.e. flow in a pipe for this study), where the flow field is predominantly shearing. This can be expected due to the presence of shear thinning effects, but the absence of any chain stretching within the GNF framework.
- The GNF-X formalism provides a way to incorporate the effects of local shear and extensional strains in the fluid viscosity. However, using the original GNF-X model, with suitable intepretations for the shear and extensional viscosities, resulted in some inconsistencies relative to the FENE-P.
- Based on our observations, we propose suitable modifications to the GNF-X formalism (termed as GNF-XM). This modified GNF-XM is able to predict trends that agree well with actual FENE-P simulations. Additionally, the GNF-XM is computationally much cheaperthan FENE-P, taking nearly an order of magnitude lesser simulation times relative to the FENE-P, especially at larger flow rates.The drag coefficient predictions from thee GNF-XM model agrees well with those from FENE-P.

Thus, this study proves the possibility of a simpler alternative, based on the GNF framework, for approximately solving flow problems involving any given constitutive equation. Here, the selected constitutive model, FENE-P, is extremely challenging to converge numerically, especially at higher flow rates. However, the GNF alternative, GNF-XM, converges much faster to provide solutions that agree well with FENE-P. A similar extension can possibly be performed for any given constitutive model, which is tough to converge numerically.

# Author Declaration

## Conflict of Interest

The authors have no conflicts to disclose.

## Data Availabilty Statement

The data that support the findings of this study are available from the corresponding author upon reasonable request.